\documentclass[10pt]{article}

\usepackage{amsmath}
\usepackage{amssymb}
\usepackage{graphics}

\title{\textbf{Probability Theories with Dynamic Causal Structure: A New Framework for Quantum Gravity}}

\author{Lucien Hardy\\
\textit{Perimeter Institute,}\\
\textit{31 Caroline Street North,}\\
\textit{Waterloo, Ontario N2L 2Y5, Canada}}

\begin{document}

\maketitle

\begin{abstract}
Quantum theory is a probabilistic theory with fixed causal structure.  General relativity is a deterministic theory
but where the causal structure is dynamic.  It is reasonable to expect that quantum gravity will be a probabilistic
theory with dynamic causal structure.  The purpose of this paper is to present a framework for such a probability
calculus.  We define an operational notion of space-time, this being composed of elementary regions.  Central to
this formalism is an object we call the {\it causaloid}. This object captures information about causal structure
implicit in the data by quantifying the way in which the number of measurements required to establish a state for a
composite region is reduced when there is a causal connection between the component regions.  This formalism puts
all elementary regions on an equal footing.  It does not require that we impose fixed causal structure. In
particular, it is not necessary to assume the existence of a background time.  The causaloid formalism does for
probability theory something analogous to what Riemannian calculus does for geometry. Remarkably, given the
causaloid, we can calculate all relevant probabilities and so the causaloid is sufficient to specify the predictive
aspect of a physical theory. We show how certain causaloids can be represented by suggestive diagrams and we show
how to represent both classical probability theory and quantum theory by a causaloid. We do not give a causaloid
formulation for general relativity though we speculate that this is possible. The causaloid formalism is likely to
be very powerful since the basic equations remain unchanged when we go between different theories - the differences
between these theories being contained in the specification of the causaloid alone. The work presented here
suggests a research program aimed at finding a theory of quantum gravity.  The idea is to use the causaloid
formalism along with principles taken from the two theories to marry the dynamic causal structure of general
relativity with the probabilistic structure of quantum theory.
\end{abstract}

\section{Introduction}

The two great pillars of twentieth century physics are general relativity (GR) and quantum theory (QT) and both
have enjoyed considerable empirical success.  It so happens that the domain where general relativity has been well
verified corresponds to situations where quantum effects (such as superposition) are negligible. And similarly, the
domain where quantum theory has been verified corresponds to situations where general relativistic effects (such as
matter dependent curvature of space time) are negligible.  Sufficiently sophisticated experiments would be able to
probe domains where both quantum and general relativistic effects are significant. However, each theory is
formulated in a way that requires that the particular effects of the other can be ignored and so, in such domains,
we would not be able to make predictions.  What is required is a new theory, a theory of quantum gravity (QG),
which reduces to GR or to QT in the situation where quantum effects or where general relativistic effects,
respectively, are small. The problem is that we want to proceed from two less fundamental theories (GR and QT) to a
more fundamental theory (QG).   How can we do this? One approach is to try to formulate one theory entirely in the
terms of the other.  For example, we might try to \lq\lq quantize general relativity". This is likely to work when
one theory is clearly less fundamental than the other. However, GR and QT each bring fundamental notions to the
table that cannot easily be accommodated in terms of the structures available in the other theory. Instead a more
even handed approach seems favourable.  This is problematic.  It seems unlikely that we can combine two
mathematical formulations of two different theories in an even handed way without stepping outside those
mathematical formulations.  Hence we adopt the following strategy.  We will pick out essential conceptual
properties of each theory and try to find a mathematical framework which can accommodate them.  A historical
example of this approach is provided by Einstein himself in his invention of special relativity which resulted from
an attempt to combine Newtonian physics with electromagnetism.  From Newtonian physics he took the Galilean
principle of invariance for inertial frames and from electromagnetism he took the fact that the speed of light is
independent of the source. These facts were set apart from their mathematical formulation in their original
theories.  Thus, Einstein stated Galileo's principle in words rather than giving it the usual mathematical
expression as the Galilean transformations.   It was only having done this that he was able to avoid the mess
associated with earlier attempts to reconcile Newtonian physics with electromagnetism in terms of the properties of
an ether.  Indeed, these earlier attempts were an attempt to formulate electromagnetism in within the Newtonian
framework.

With the implementation of this approach in mind, we note the following.
\begin{enumerate}
\item General relativity is a deterministic theory with dynamic causal structure.
\item Quantum theory is a probabilistic theory with fixed causal structure.
\end{enumerate}
Once the probabilistic cat is out of the bag it is unlikely that we will go back to a fundamentally deterministic
theory.  Likewise, once we have dynamic causal structure it is unlikely that a more fundamental theory will have an
underlying fixed causal structure.  Hence, we require a mathematical framework for physical theories with the
following properties:
\begin{enumerate}
\item It is probabilistic.
\item It admits dynamic causal structure.
\end{enumerate}
In this paper we will find such a framework.  We will show how QT can be formulated in this framework.  We also
expect to be able to formulate GR in the framework though we do not give an explicit construction.  But, of course,
the real point of this exercise is that we should be able to formulate a theory of QG in this framework. And,
further, this framework should make this job easier.  We will suggest possible approaches to finding a theory of QG
within this framework.

\section{Overview}

In GR we introduce coordinates $x^\mu$.  We can consider intervals $\delta x^\mu$ in these coordinates.  We so not
say up front which of these intervals (or which linear combinations of these intervals) is time-like.   It is only
after solving for the metric that we can do this.   The causal structure is dynamic.   In quantum theory, on the
other hand, we must specify the causal structure in advance.  One way to see this is to consider different ways in
which we might put two operators, $\hat A$ and $\hat B$, together.  If the two regions corresponding to these
operators are space-like separated then we use the tensor product $\hat A\otimes \hat B$.   If the two regions are
immediately sequential (time-like) then we write $\hat B \hat A$.  In order to know what type of product to take we
need to know the causal structure in advance.  We seek a new type of product which unifies these two products
(along with any other products in QT) and puts them on an equal footing.

The approach taken in this paper is operational.  We define an operational notion of space-time consisting of
elementary regions $R_x$.  An arbitrary region $R_1$ may consist of many elementary regions.   In region $R_1$ we
may perform some action which we denote by $F_{R_1}$ (for example we may set a Stern-Gerlach apparatus to measure
spin along a certain direction) and observe something $X_{R_1}$ (the outcome of the spin measurement for example).
Consider two disjoint regions $R_1$ and $R_2$. Our basic objective is to find a formalism which allows us to
calculate the probability for something in one region, $R_1$, conditioned on what happened in another region $R_2$
if this probability is well defined (we will explain what we mean by a \lq\lq well defined probability" in Sec.\
\ref{statement}). Namely we want to be able to calculate all probabilities of the form
\begin{equation}
{\rm prob}(X_{R_1}|F_{R_1}, X_{R_2}, F_{R_2})
\end{equation}
when well defined.   We would like the formalism that does this to put every elementary region $R_x$ on an equal
footing.

To this end we introduce vectors ${\bf r}_{(X_{R_1}, F_{R_1})}(R_1)$ for $R_1$ (these are analogous to operators in
QT) . Such vectors are defined for any region including the elementary regions $R_x$.  Given any composite region
such as $R_1\cup R_2$ we can find the corresponding ${\bf r}$ vector using the {\it causaloid product}
\begin{equation}
{\bf r}_{(X_{R_1}\cup X_{R_2}, F_{R_1}\cup F_{R_2})}(R_1\cup R_2) = {\bf r}_{(X_{R_1},
F_{R_1})}(R_1)\otimes^\Lambda {\bf r}_{(X_{R_2}, F_{R_2})}(R_2)
\end{equation}
This means that ${\bf r}$ vectors for any region can be built out of ${\bf r}$ vectors for the elementary regions,
$R_x$, comprising this region.   The causaloid product is given by the {\it causaloid} which we will describe
briefly in a moment.  The  ${\bf r}$ vectors for the elementary regions themselves are also given by the causaloid.
If this formalism is applied to QT then the causaloid product unifies the different products in QT mentioned above.

We find that the probability
\[{\rm prob}(X_{R_1}|X_{R_2}, F_{R_1}, F_{R_2})   \]
is well defined if and only if
\[ {\bf v}\equiv {\bf r}_{(X_{R_1},F_{R_1})}\otimes^{\Lambda} {\bf r}_{(X_{R_2}, F_{R_2})}  \]
is parallel to
\[ {\bf u}\equiv \sum_{ Y_{R_1}}{\bf r}_{(Y_{R_1},F_{R_1})}\otimes^{\Lambda} {\bf r}_{(X_{R_2}, F_{R_2})}  \]
(where the sum is over all possible observations, $Y_{R_1}$, in $R_1$ consistent with action $F_{R_1}$) and this
probability is given by
\begin{equation}
{\rm prob}(X_{R_1}|X_{R_2}, F_{R_1}, F_{R_2})  = \frac{|{\bf v}|}{|{\bf u}|}
\end{equation}
where $|{\bf a}|$ denotes the length of the vector ${\bf a}$.

The causaloid is theory specific and is given by providing a means to calculate certain matrices (called lambda
matrices). The lambda matrices quantify the way in which the number of measurements to determine the state is
reduced due to correlations implied by the theory. We have lambda matrices for each elementary region (called local
lambda matrices) and we have lambda matrices for every subset of elementary regions.  Hence, at this general level
all elementary regions are put on an equal footing.   In any specific theory we will expect that some lambda
matrices will follow from others. In QT, for example, it turns out that we only need local lambda matrices and
lambda matrices for pairs of adjacent regions. From these we can calculate all other lambda matrices. In a
particular theory we will expect to break the symmetry between elementary regions by virtue of some particular
choice of lambda matrices. For example, in QT the lambda matrix associated with a pair of adjacent elementary
region is different to the lambda matrix associated with a pair of non-adjacent elementary regions. However, the
fact that we start with a formalism that does not impose any particular such structure from the very beginning puts
us in a strong position to make progress in finding a theory of QG.

The causaloid framework does not have, as a fundamental notion, the idea of a state evolving in time.  However,
standard physical theories such as QT do.  Thus, to help us put the QT in the causaloid framework, we will show how
to recover a notion of a state evolving in time in the causaloid framework.  Having done this we are able to put
classical probability theory and quantum theory into the causaloid framework.  Having pulled QT into the causaloid
framework we can leave behind the problematic notion of a state at time $t$.

Any attempt to find a theory of QG in this program is likely to start by putting GR into the framework.  We discuss
how this might be done before considering issues that arise in QG.

The important new technical results in this paper are contained in Sec.\ \ref{meatstarts} to Sec.\ \ref{meatends}.
These sections are fairly self-contained and the impatient reader can jump straight to those sections on a first
reading of this paper (though perhaps skimming the earlier sections).

\section{Data}\label{Datumism}

We are looking for a framework for physical theories.  But what is a physical theory and what does it do for us?
There are many possible answers to these questions.  But we take the following to be true:
\begin{quote}
{\bf Assertion:}  A physical theory, whatever else it does, must correlate recorded data.
\end{quote}
A physical theory may do much more than this. For example it may provide a picture of reality.  It may satisfy our
need for explanation by being based on some simple principles. But for a physical theory to have any empirical
content, it must at least correlate recorded data.  This sounds like a rather weak assertion. But, as we will see,
it provides us with a strong starting point for the construction of the framework we seek.

The assertion above leaves unspecified what the word \lq\lq correlate" means.  This could be deterministic
correlation, probabilistic correlation, or conceivably something else.  Since we will be interested in
probabilistic theories, we will take this to mean probabilistic correlation.

What is data?   We can compile the following list of properties that data has.
\begin{enumerate}
\item Data is a record of (i) actions and (ii) observations.  For example it might record statements like (i) I
lifted the rock and let go (an action), and (ii) it fell and hit my toe (an observation).
\item Data is recorded by physical means. For example it may be written on bits of paper, stored in a computer's
memory, or stored in the brain of the experimentalist.
\item Data is robust (it is unlikely to randomly change).
\item Data can be copied so that new physical records exist.
\item Data can be translated (e.g. English to French or binary to base 10).
\item Data can be moved around (e.g. in wires or on bits of paper).
\item Data can be processed (e.g. to check that it is correlated according to some physical theory).
\end{enumerate}
The physicality of data may concern us a little - especially in those situations where we expect the physical
systems which store, transport, and process the data to interfere with the physical experiment we are performing.
To deal with this concern we make the following assumption
\begin{quote}
{\bf The indifference to data principle}:  It is always possible to find physical devices capable of storing,
transporting, and processing data such that (to within some arbitrarily small error) the probabilities obtained in
an experiment do not depend  on the detailed configuration of these devices where this detailed configuration
corresponds to the particular data and programming (for the program which will process the data) whilst it is being
stored, transported, and processed.  Such physical devices will be called {\it low key}.
\end{quote}
Without such a principle the probabilities might depend on whether the experiment is conducted by an Englishman or
a Frenchman (since the same data in English or French will have a different detailed physical configuration). This
principle does not imply that the presence of the physical device which stores and processes the data has no effect
on the experiment. But rather that any such effect does not depend on the detail of the data being stored. For
example, a computer being used to record and process data from a nearby gravitationally sensitive experiment has
mass and therefore will effect the experiment in question. However, this effect will not depend on the detailed
configuration of the computer which corresponds to the data (or at least that effect will be arbitrarily small).
Consequently the principle does not forbid the physical data devices from being part of the experiment. For
example, we could throw a computer from the leaning tower of Pisa to gain information about how the computer falls.
It might collect data through a camera about the time it passes successive levels of the building.  In this case
the data device is actually part of the experiment and the principle still applies. This means that we do not need
to put the observer (for observer read \lq\lq physical data devices") outside the system under investigation. The
observer can be a part of the system they are investigating so long as they can store and process data in a low key
manner. In fact, one might even argue that we must always regard the observer as part of the system we are
observing. How could the data end up being collected otherwise?  Of course, there are certain situations where we
have reasons to regard the observer as being outside the system under consideration - namely those situations where
the probabilities measured do not depend on the bulk properties of the observer. However, the important point is
that we can do physics when this is not the case so long as we can have low key data devices.   It is easy to
imagine data processing devices which are not low key. For example, we could use a computer which stores
information in the configuration of a number of large rocks rather than a standard electronic computer to store and
process data about about a nearby gravitationally sensitive experiment. Then the probabilities would depend on the
detail of the data.

The fact that we start with considerations about data where data is a collections of actions and observations puts
us in an operational or instrumental mode of thinking.  Operationalism played a big role in the discovery of both
relativity theory and QT.  There are different ways of thinking about operationalism. We can either take it to be
fundamental and assert that physical theories are about the behaviour of instruments and nothing more.  Or we can
take it to be a methodology aimed at finding a theory in which the fundamental entities are beyond the operational
realm. In the latter case operationalism helps us put in place a scaffolding from which we can attempt to construct
the fundamental theory. Once the scaffolding has served its purpose it can be removed leaving the fundamental
theory partially or fully constructed. The physicist operates best as a philosophical opportunist (and indeed as a
mathematical opportunist). For this reason we will not commit to either point of view for the time being noting
only that the methodology of operationalism serves our purposes. Indeed, operationalism is an important weapon in
our armory when we are faced with trying to reconcile apparently irreconcilable theories. A likely reason for any
such apparent irreconcilability is that we are making some unwarranted assumptions beyond the operational realm. It
was through careful operational reasoning that Einstein was able to see that absolute simultaneity is unnecessary
(since it has no operational counterpart).  The operational methodology is a way of not making wrong statements. If
we are lucky we can use it to make progress.

\section{Remarks on quantum theory}

When all is said and done, quantum theory provides a way of calculating probabilities.  It is a probability
calculus.   Hence, its natural predecessor is not Newtonian mechanics or any other branch of classical physics, but
rather what might be called classical probability theory (CProbT).   Thus, in the same way that CProbT can be
applied to various physical situations from classical physics (such as systems of interacting spins, particles in a
gas, electromagnetic fields...) to calculate probabilities,  quantum theory can be applied to various different
physical situations (interacting quantum spins,  a quantum particle in a potential well, quantum fields, ....) to
calculate probabilities. Quantum theory is, like classical probability theory, a meta theory with many realizations
for different physical situations.

One particular realization of QT is what might be called quantum mechanics (QM).  This is the non-relativistic
theory for multi-particle systems in which we introduce a wavefunction $\psi(x_1,x_2,\cdots, t)$ and the
Schr\"odinger equation
\begin{equation}
i\hbar \frac{\partial \psi}{\partial t} = - \sum_i \frac{\hbar^2}{2m_i} \nabla_i^2 \psi + V(x_1,x_2\cdots, t) \psi
\end{equation}
to evolve the wavefunction.  QM is an example of QT.

Quantum field theory (QFT) is another example of QT. The basic framework of quantum theory (which we will present
in detail below) consisting of an evolving state $\hat\rho$ that acts on a Hilbert space $\cal H$ is capable of
expressing both non-relativistic quantum mechanics and relativistic QFT (see pg.\ 49 of \cite{Weinberg}).  This is
clearest in the formulation of QFT in which we write down a superwavefunctional $\Psi(\phi(x))$ which we can regard
as a linear superposition of basis states where these basis states correspond to definite configurations of the
field $\phi(x)$.

This addresses a common misconception.  We should regard QFT as a special case of QT rather than something like the
converse.  Thus we should not think of QT as a limiting case of QFT  - though we might attempt to derive QM as the
limit of QFT. It is not the case that QT, thus understood, is necessarily non-relativistic. The only point that
should be added to these remarks is that QFT requires an infinite dimensional Hilbert space whereas we can do a lot
in non-relativistic scenarios with finite dimensional Hilbert spaces. However, this is a technical rather than
conceptual point and, in any case, there are good reasons to believe that a theory of quantum gravity will have
something like a finite dimensional Hilbert space. In our discussion of QT we will stick with finite dimensional
Hilbert spaces both for technical simplicity and because we are dealing with finite data sets. Issues related to
infinities and continuities will be discussed in  Sec.\ \ref{newcalculus}.

In going from QT to QFT we have to add much additional structure to QT.  However, the deep conceptual novelties of
QT are evident without going to QFT.  For this reason it seems reasonable to look at ways to combine QT (rather
than QFT) with GR.  This way we can hope to import these conceptual novelties into a theory of QG without getting
distracted by the additional structure of QFT.  Ultimately we would require that a theory of QG incorporate QFT (at
some appropriate level of approximation).  However, we take the attitude that this is not likely to be important in
the early days of the construction of QG and it may even be possible to fully construct QG before taking on this
consideration.

\section{Basic framework for operational theories}

We want to give a simple operational formulation of CProbT and QT and for this purpose we present a framework which
works for probabilistic theories that admit a universal background time.

\begin{figure*}[t]
\resizebox{\textwidth}{!} {\includegraphics{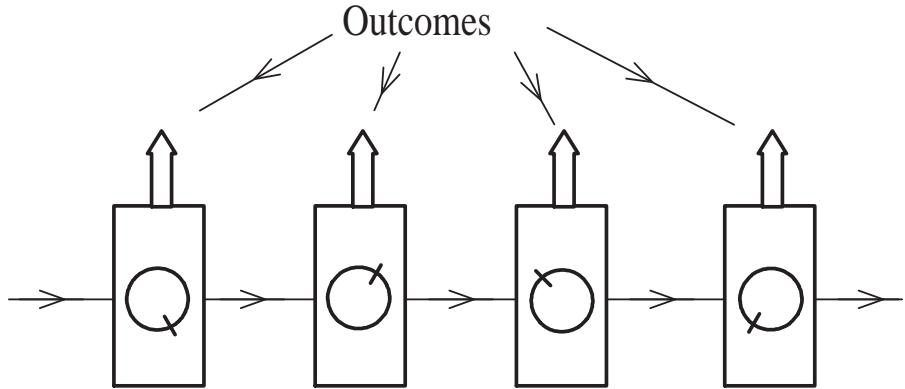}} \caption{\small Sequence of operations acting on
system.  Each operation has a knob to vary the operation enacted and an outcome which is recorded as
data.}\label{basicscenario}
\end{figure*}

The basic scenario we consider is that shown in Fig.\ \ref{basicscenario}.   This consists of a sequence of
operations on the system. We can represent these operations by boxes.  Each box has a knob on it which can be used
to vary the operation implemented.  At each operation we have the possibility of extracting some outcome $l$ (this
is data). Each operation can be regarded alternatively as (a) a preparation (since it outputs a system), (b) a
transformation (since it transforms the state of a system),  and (c) a measurement (since it inputs a system and
outputs an outcome $l$). The same can be said of any sequence of such operations. We can, strictly, only regard an
operation as a preparation if it outputs the system in a definite state.  We define the state in the following way.
\begin{quote}
{\bf The state} associated with a preparation is that thing represented by any mathematical object that can be used
to calculate the probability for every outcome of every measurement that may be performed on the system.
\end{quote}
Given this definition we can define the state to be represented by a list of all probabilities,
\begin{equation}
{\bf P} = \left( \begin{array}{c}  \vdots \\ p_{\alpha} \\ \vdots  \end{array} \right)
\end{equation}
where $\alpha$ labels every outcome of every possible measurement.    We note that we can write
\begin{equation}\label{lastresort}
p_\alpha = {\bf R}_\alpha \cdot {\bf P}
\end{equation}
where ${\bf R}_\alpha$ has a 1 in position $\alpha$ and 0's everywhere else. The object ${\bf P}$ contains a lot
of information.  In general we would expect a physical theory to afford some simplification so that some entries
in ${\bf P}$ can be calculated from other entries.  In fact we can insist that this be done by a linear formula so
we have a state given by
\begin{equation}
{\bf p} = \left( \begin{array}{c}  p_1 \\ p_2 \\ \vdots \\ p_K \end{array} \right)
\end{equation}
such that
\begin{equation}
p_\alpha = {\bf r}_\alpha \cdot {\bf p}
\end{equation}
The $p_k$'s are the probabilities associated with a set of {\it fiducial} measurement outcomes.  We take $K$ to be
the minimum number of entries in ${\bf p}$ that makes it possible to write the state by a linear formula like this.
That this will always be possible is clear since we have (\ref{lastresort}) as a last resort.

A special measurement is the identity measurement ${\bf r}^I$ corresponding to the measurement whose result is
positive if any result is seen. In the case that the state is normalized we have ${\bf r}^I \cdot{\bf p} = 1$.
However, for technical reasons, we will not normalize the state after each step.

It follows from the fact that these probabilities are given by a linear formula that the transformation of the
state is given by a linear formula.  Thus, if we obtain result $l$ the new state is
\begin{equation}
{\bf p} \rightarrow Z_l {\bf p}
\end{equation}
where $Z_l$ is a $K\times K$ real matrix.   This is clear since each component of the new state must be given by a
linear combination of the components in the initial state.  The probability of outcome $l$ is
\begin{equation}
{\rm prob}_l= \frac{{\bf r}_I \cdot Z_l {\bf p}}{{\bf r}_I \cdot {\bf p}}
\end{equation}
The state can be normalized by dividing by ${\bf r}^I\cdot {\bf p}$.  However, this introduces unnecessary
non-linearities in the evolution of the state.   It is more convenient to allow the state to be unnormalized and
use the above formula for calculating probabilities.

We may have more than one system.  We need a way of setting up the framework for such composite systems.  Both
CProbT and QT turn out to be simple in this respect.   The state of a composite system is given by specifying joint
probabilities $p_{k_1k_2}$ with $k_1=1$ to $K_1$ and $k_2=1$ to $K_2$.

\section{Brief summary of classical probability theory}\label{CProbT}

Consider a classical system which can be in one of $N$ distinguishable configurations (for example a bit has
$N=2$). We can write the state of this system by specifying a probability, $p_n$, for each configuration, $n$.
\begin{equation}
{\bf p} = \left( \begin{array}{c}  p_1 \\ p_2 \\ \vdots \\ p_N \end{array} \right)
\end{equation}
Note that $K=N$. We define the {\it identity measurement vector}, ${\bf r}_I$, for CProbT by
\begin{equation}
{\bf r}_I = \left( \begin{array}{c}  1 \\ 1 \\ \vdots \\ 1 \end{array} \right)
\end{equation}
Now we can state postulates of CProbT in compact form.
\begin{enumerate}
\item  The state of a system is given by ${\bf p}\in S_N$ where $S_N$ is defined by (i) $p_n\geq 0$ and (ii)
${\bf r}_I\cdot {\bf p} \leq 1$.
\item The state ${\bf p}$ for a composite system $12$ made from systems $1$ and $2$ has components $p_{n_1n_2}$ and
belongs to $S_{N_1N_2}$.
\item  Any operation which transforms of the state of a system and has classical outcomes labeled by $l$
is associated with a set of $N\times N$ matrices $Z_l$ which (i) map $S_N$ into $S_N$, and (ii) have the
properties that ${\bf r}_I\cdot Z_l {\bf p} \leq {\bf r}_I\cdot {\bf p}$ and and ${\bf r}_I\cdot (\sum_l Z_l) {\bf
p}= {\bf r}_I\cdot {\bf p}$ for all states. The probability of outcome $l$ is
\begin{equation}\label{clasprob}
{\rm prob}_l= \frac{{\bf r}_I \cdot Z_l {\bf p}}{{\bf r}_I \cdot {\bf p}}
\end{equation}
and the state after outcome $l$ is observed is
\begin{equation}
{\bf p} \rightarrow Z_l {\bf p}
\end{equation}
\end{enumerate}
A few notes of clarification are useful here.  We deliberately do not impose the normalization condition ${\bf
r}_I\cdot {\bf p} = 1$ (though we impose the condition ${\bf r}_I\cdot {\bf p} \leq 1$ to keep these vectors
bounded). First it is not necessary to normalize since the denominator on the RHS of (\ref{clasprob}) ensures that
the sum of probabilities over all outcomes adds up to 1. Second, it is useful to allow the freedom not to normalize
since then we can regard $Z_l{\bf p}$ as a new state even though this new state is not normalized.  We can, if we
wish, normalize a state by dividing it by ${\bf r}_I\cdot {\bf p}$.

\section{Brief summary of quantum theory}

The postulates of quantum theory (stripped of additional structure pertaining to particular applications) can be
written in the following compact form.
\begin{enumerate}
\item The state of a system is given by a positive operator $\hat{\rho}$ acting on a complex Hilbert space ${\cal H}$
with ${\rm trace}(\hat\rho)\leq 1$.
\item A composite system $12$ made from systems $1$ and $2$ has Hilbert space ${\cal H}_1
\otimes {\cal H}_2$.
\item Any operation which transforms of the state of a system and has classical outcomes labelled by $l$
is associated with a set of trace non-increasing completely positive linear maps (also known as superoperators),
$\$_l( \cdot)$, where $\sum_l \$_l $ is trace preserving. The probability of outcome $l$ is
\begin{equation}
{\rm prob}_l= \frac {{\rm trace}(\$_l( \hat{\rho}))} {{\rm trace}(\hat{\rho})}
\end{equation}
and the state after outcome $l$ is observed is
\begin{equation}
\hat{\rho} \rightarrow \$_l (\hat{\rho})
\end{equation}
\end{enumerate}
A completely positive linear map $\$ $ is one which, when extended to a composite system $12$ as $ \$ \otimes I$
(where $I$ is the identity map on system $2$), leaves the state for the total system $\hat{\rho}_{12}$ positive
regardless of the initial state of the total system and the dimension of system 2. This property is required for
the internal consistency of the theory. There are two familiar special cases of superoperators. First there are
unitary maps,
\begin{equation}
\$ (\hat{\rho}) = \hat{U} \rho \hat{U}^\dagger
\end{equation}
where $\hat{U}$ is a unitary operator (satisfying $\hat{U}\hat{U}^\dagger=I$).  We can understand this as an
example of postulate 2 above where $l$ only takes one value (so we always see the same result).  Second there are
projection operators,
\begin{equation}
\$ (\rho) = \hat{P} \rho \hat{P}
\end{equation}
where $\hat{P}$ is a projection operator (satisfying $\hat{P}\hat{P}=\hat{P}$).  The first example is trace
preserving and the second example is trace decreasing.  There are many other types of superoperator which, in one
way or another, extrapolate between these two extremes.

States can normalized at any time by applying the following formula
\begin{equation}
\hat{\rho} \rightarrow \frac{1}{{\rm trace}(\hat{\rho})} \hat{\rho}
\end{equation}
In calculating the probabilities for outcomes of measurements we are not necessarily interested in the evolution of
the state of the system during measurement.   To this end we can associate a set, $\{\hat A_l \}$, of positive
operators with
\begin{equation}\label{Arho}
{\rm prob}_l = {\rm trace}( \hat{A}_l \hat{\rho}) = {\rm trace}( \$_l(\hat{\rho}))
\end{equation}
if the state is normalized.   There is a many to one linear map between $\$_l $ and $\hat{A}_l$ (which can be
written down explicitly).  Usually positive operators $\hat{A}$ (or something like them) are introduced in the
postulates. However, this is not necessary.

\section{Quantum theory formulated in similar way to classical probability theory}\label{QTrZp}

We can reformulate QT in a way which resembles CProbT.  We define
\begin{equation}
\widehat{\bf A} = \left( \begin{array}{c}  \widehat{A}_1 \\ \widehat{A}_2 \\ \vdots \\ \widehat{A}_{N^2}
\end{array} \right)
\end{equation}
where $\widehat{A}_k$ for $k=1$ to $N^2$ are a fixed fiducial set of linearly independent positive operators which
span the space of positive operators acting on an $N$ dimensional Hilbert space ${\cal H}_N$.   We define the {\it
identity measurement vector}, ${\bf r}_I$, for QT by
\begin{equation}
{\bf r}_I \cdot \widehat{\bf A} = I
\end{equation}
where $I$ is the $N^2\times N^2$ identity matrix.  Since the operators $\widehat{A}_k$ form a complete linearly
independent set, ${\bf r}_I$ is unique.   Now we can restate the postulates of QT.
\begin{enumerate}
\item  The state of a system is given by ${\bf p}\in S^Q_N$ where $S^Q_N$ is the set of ${\bf p}$ for which (i) we can
write ${\bf p}={\rm trace}(\widehat{\bf A} \hat \rho)$ where $\hat\rho$ is a positive operator acting on ${\cal
H}_N$ and (ii) ${\bf r}_I\cdot {\bf p} \leq 1$.
\item The states for a composite system $12$ made from systems $1$ and $2$ belong to
$S^Q_{N_1N_2}$ derived from $\widehat{\bf A}_1\otimes\widehat{\bf A}_2$ whose elements act on ${\cal
H}_1\otimes{\cal H}_2$.
\item  Any operation which transforms of the state of a system and has classical outcomes labelled by $l$
is associated with a set $N^2\times N^2$ matrices $Z_l$ which (i) are such that $Z_l\otimes I$ maps $S^Q_{NM}$ into
$S^Q_{NM}$ for any ancillary system of any $M$, and (ii) have the property that ${\bf r}_I\cdot Z_l {\bf p} \leq
{\bf r}_I\cdot {\bf p}$ and also ${\bf r}_I\cdot (\sum_l Z_l) {\bf p}= {\bf r}_I\cdot {\bf p}$ for all states. The
probability of outcome $l$ is
\begin{equation}
{\rm prob}_l= \frac{{\bf r}_I \cdot Z_l {\bf p}}{{\bf r}_I \cdot {\bf p}}
\end{equation}
and the state after outcome $l$ is observed is
\begin{equation}
{\bf p} \rightarrow Z_l {\bf p}
\end{equation}
\end{enumerate}
Note that for QT we have $K=N^2$.  It is easy to show that
\begin{equation}\label{dollarZ}
Z = {\rm trace}(\widehat{\bf A}\$ (\widehat{\bf A}^T)) [{\rm trace}(\widehat{\bf A} \widehat{\bf A}^T)]^{-1}
\end{equation}
where the superscript $T$ denotes transpose.

\section{Reasonable postulates for quantum theory}\label{QTpostulates}

The postulates of QT, unlike those of CProbT, are rather abstract. It is not clear where they come from.  However,
QT can be obtained by a set of what might be regarded as reasonable postulates \cite{hardy}.  Before stating these
we need a few definitions.
\begin{quote}
{\bf The maximum number of reliably distinguishable states}, N, is defined to be equal to the maximum number of
members of any set of states which have the property that there exists some measurement device which can be used
to distinguish the states in a single shot measurement (so that the sets of outcomes possible for each state in
this set are disjoint).
\end{quote}
This quantity captures the {\it information carrying capacity} of the system (in QT it is simply the dimension of
the Hilbert space).  More exactly, we might say that the information carrying capacity of the system is $\log_2 N$
bits. We will say that
\begin{quote}
{\bf A system is constrained to have information carrying capacity} $\log_2 M$ if the states are such that, for a
measurement set to distinguish $N$ distinguishable states, we only ever obtain outcomes associated with some given
subset of $M$ of these states.
\end{quote}
We want to define a useful notion
\begin{quote}
{\bf Equivalence classes}.  Two operations belong to the same equivalence class if replacing one by the other gives
rise to the same probabilities
\end{quote}
An example might be measuring the polarization of a photon with a polarizing beamsplitter or a calcite crystal each
orientated at an angle $\theta$.  If one device is replaced by the other (with an appropriate identification of
outcomes between the two devices) then the probabilities remain the same.

The spin degree of freedom of an electron and the polarization degree of freedom of a photon are, in each case,
described by a Hilbert space of dimension 2.  There is a sense in which these two systems have the same properties.
We define this idea as follows
\begin{quote}
{\bf Two systems have the same properties} if there is a mapping between equivalence classes of operations such
that under this mapping we get the same probabilities for outcomes for each type of system.
\end{quote}
This mapping might take us from an experiment involving an electron's spin degree of freedom to another experiment
involving a photon's polarization degree of freedom.  For example, the electron could be prepared with spin up
along the $z$ direction, sent through a magnetic field which (acting as a transformation) rotates the spin through
$20^\circ$ in the $zx$ plane.  Then the electron impinges on a Stern-Gerlach device orientated in the $x$
direction. Under a mapping between equivalence classes, this might correspond to an experiment in which a photon is
prepared with vertical polarization.  The photon passes through a crystal which rotates its polarization through
$10^\circ$ and then onto a polarizing beamsplitter orientated at angle $45^\circ$.  The probabilities seen would be
the same in each case (note that the angles must be halved for the photon since orthogonal states are vertical and
horizontal rather than up and down).

We define
\begin{quote}
{\bf Pure states} are states which cannot be simulated by mixtures of other distinct states
\end{quote}
Also we define
\begin{quote}
{\bf A reversible transformation} is a transformation, $T$, for which there exists another transformation $T^{-1}$
which is such that if $T$ is applied, then $T^{-1}$, the over all transformation leaves the the incoming state
unchanged (it is the identity transformation) for all incoming states
\end{quote}
and
\begin{quote}
{\bf A continuous transformation} is one which can be enacted by sequential application of infinitely many
infinitesimal transformations where an infinitesimal transformation is one which has the property that it only has
an infinitesimal effect on the probability associated with any given outcome for any measurement that may be
performed on the state
\end{quote}

Given these definitions, we are in a position to state the postulates
\begin{description}
\item[Postulate 0] {\it Probabilities}.  Relative frequencies (measured by
taking the proportion of times a particular outcome is observed) tend to the same value (which we call the
probability) for any case where a given measurement is performed on a ensemble of $n$ systems prepared by some
given preparation in the limit as $n$ becomes infinite.
\item[Postulate 1] {\it Information}. There exist systems for which
$N=1,2,\cdots$, and, furthermore, systems having, or constrained to have the same information carrying capacity
have the same properties.
\item[Postulate 2]  {\it Composite systems}. A composite system $12$ consisting of subsystems $1$
and $2$ satisfies $N_{12}=N_1N_2$ and $K_{12}=K_1K_2$.
\item[Postulate 3] {\it Continuity}. There exists a continuous reversible
transformation on a system between any two pure states of that system for systems of any dimension $N$.
\item[Postulate 4] {\it Simplicity}. For each given $N$, $K$ takes the
minimum value consistent with the other axioms.
\end{description}
The crucial postulate here is Postulate 3. If the single word \lq\lq continuous" is dropped from these axioms, then
we get classical probability theory rather than quantum theory.

These axioms give rise to the full structure of quantum theory with operators on finite dimensional complex Hilbert
space as described above.   The construction theorem is simple but rather lengthy and the reader is referred to
\cite{hardy} for details.  However, the main ideas are as follows.  If follows from Postulate 1 that $K=K(N)$ and
$K(N+1)>K(N)$ (this second point is not obvious).   It then follows from Postulate 2, after a little number theory,
that $K=N^r$ where $r=1,2,3,\dots$.  By the simplicity postulate we should take the smallest value of $r$ that
works. First we try $r=1$ but this fails because of the continuity postulate.  Then we try $r=2$ and this works.
Thus, we have $K=N^2$. (As an aside, if we dropped the word \lq\lq continuity" we get $r=1$ and hence $K=N$. This
leads very quickly to classical probability theory.)  Next we take the simplest nontrivial case $N=2$, and $K=4$.
If we just consider normalized states then rather than 4 probabilities we have three.  We apply the group of
continuous reversible transformations (implied by the continuity postulate) to show that the set of states must
live inside a ball (with pure states on the surface). This is the Bloch ball of quantum theory for a two
dimensional Hilbert space. Thus, we get the correct space of states for two dimensional Hilbert space. We now apply
the information postulate to the general $N$ case to impose that the state restricted to any two dimensional
Hilbert space behaves as a state for Hilbert space of dimension 2.  By this method we can construct the space of
states for general $N$. Various considerations give us the correct space of measurements and transformations and
the tensor product rule and, thereby, we reconstruct quantum theory for finite $N$.

\section{Remarks on general relativity}\label{remarksGR}

General relativity was a result of yet another attempt to make two theories consistent, namely special relativity
and Newton's theory of gravitation.  Einstein gives various reasons that Galileo's principle of invariance is not
sufficiently general since it applies only to inertial frames.    He replaces it with the following
\begin{quote}
The general laws of nature are to be expressed by equations which hold good for all systems of co-ordinates, that
is, are co-variant with respect to any substitutions whatever (generally covariant) \cite{Einstein}.
\end{quote}
He then employs the equivalence principle to argue that the metric $g_{\mu\nu}$ represents the gravitational field.
He sets up the mathematics of tensors as objects which have the property that a physical law  expressible by
setting all the components of a tensor equal to zero is generally covariant.   Out of these ideas he is able, by an
ingenious chain of reasoning, to obtain field equations for GR which can be expressed as
\begin{equation}
G_{\mu\nu} = 8\pi T_{\mu\nu}
\end{equation}
$G_{\mu\nu}$ is Einstein tensor and is a measure of the local curvature concocted out of derivatives of the metric.
$T_{\mu\nu}$ is the stress-energy tensor and is determined by the local properties of matter.

We can distinguish two roles for the metric.
\begin{enumerate}
\item It determines the set of local frames in which gravitational fields vanish and inertial physics
applies locally.  This property is given by the local value of the metric without taking into account its
variation with $x^\mu$.
\item It tells us how to compare local fields at two different points by providing a way to parallel transport such
fields.  This it does via the connection $\Gamma^\alpha_{\mu\nu}$ and the machinery of covariant differentiation.
This property does depend on the variation of the metric with $x^\mu$.
\end{enumerate}
The metric also determines causal structure.   It tells us whether two events are space-like or time-like and
thereby determines whether we can send a signal from one to the other.   However, the metric is a dynamical feature
of the theory.  It is determined by solving the Einstein field equations.   Hence the causal structure is
dynamical.  We can try to express this in more operational terms.  Given local events with local labels $x^\mu$
(which may be abstract or read off some real physical reference frame) there is no way in general to say, in
advance (that is without solving the equations), whether it is possible to send a signal from one to another.  In
non-gravitational physics we have fixed causal structure.  For example, in SR the metric is fixed and so we know
the causal structure in advance.

GR is a deterministic theory.  In classical physics we can always introduce probabilities simply by applying
CProbT.  However, CProbT assumes a fixed causal structure just as QT does.  In particular it assumes a fixed
background time.  Hence, we would not expect to be able to apply CProbT to GR in a straight forward manner.
Indeed, we could consider the program of unifying CProbT and GR to find what we might call probabilistic general
relativity (ProbGR).  This program  might be expected share many of the same difficulties we face in the program
to find a theory of QG. We have taken as our goal to find a framework for probabilistic theories which admit
dynamic causal structure. This framework should include ProbGR as well as QG. Hence, we will need to introduce
further principles to get QG rather than ProbGR. For such principles we can look to the differences between CProbT
and QT.  These differences are especially clear in the formulation of QT in Sec.\ \ref{QTrZp} which looks similar
to CProbT and in the postulates introduced in Sec.\ \ref{QTpostulates} (see \cite{Hardywheelerpaper} for a
discussion of the differences between CProbT and QT).

\section{Remarks on the problem of finding a theory of quantum gravity}\label{QGsec}

The most obvious issue that arises when attempting to combine QT with GR is that QT has a state on a space-like
surface that evolves with respect to an external time whereas in GR time is part of a four dimensional manifold
whose behavior is dynamically determined by Einstein's field equations. We can formulate GR in terms of a state
evolving in time - namely a canonical formulation \cite{ADM, Ashtakar}. Such formulations are rather messy (having
to deal with the fact that time is not a non-dynamical external variable) and break the elegance of Einstein's
manifestly covariant formulation. Given that Einstein's chain of reasoning depended crucially on treating all four
space-time coordinates on an equal footing it is likely to be at least difficult to construct QG if we make the
move of going from a four dimensional manifold $\cal M$ to an artificial splitting into a three dimensional spatial
manifold $\Sigma$ and a one dimensional time manifold $R$. But there is a further reason coming from quantum theory
that suggests it may be impossible.  If the causal structure is dynamically determined then what constitutes a
space-like surface must also be dynamically determined. However, in quantum theory we expect any dynamics to be
subject to quantum uncertainty.  Hence, we would expect the property of whether a surface is space-like or not to
be subject to uncertainty.  It is not just that we must treat space and time on an equal footing but also that
there may not even be a matter-of-fact as to what is space and what is time even after we have solved the equations
(see \cite{Ishamtime, UnruhWald}).  To this end we will give a framework (which admits a formulation of quantum
theory) which does not take as fundamental the notion of an evolving state. The framework will, though, allow us to
construct states evolving through a sequence of surfaces. However, these surfaces need not be space-like (indeed,
there may not even be a useful notion of space-like).

Another way of thinking about these issues is by considering how we might probe causal structure.  The most obvious
way to do this is to use light signals since they probe the light cone structure which underpins causal structure
in GR.   In GR we are typically interested in cases where the presence of light represents only a small
perturbation and so we freely employ counterfactual reasoning in which we imagine sending or not sending light
signals without having to consider the effect this has on the solution to Einstein's field equations.  On the other
hand, in QT, the presence or not of even a single photon can have a dramatic effect on what is seen.  The most
clear example of this is provided in \cite{Mandle} in which an interference effect involving one photon depends on
whether the path of another photon is blocked or not.  In QT we have a fixed background causal structure (which
might be implemented for example by bolting apparatuses down to a rigid structure) and so there is no need to
employ such reasoning about the counterfactual transmission of photons for the purposes of understanding causal
structure. However, in QG, we will not assume that there is a fixed background causal structure.  We cannot assume
that two regions of space-time have a certain causal relationship in the absence of any photon being transmitted
between them just because we know that their causal relationship would be fixed if a photon were to be so
transmitted. This line of thinking lends separate support to the possibility mentioned above that there may not
even be a matter-of-fact as to what is space and what is time even after the equations have been solved.

A more mathematical handle on this issue can be gained by considering the various ways in quantum theory we can put
together pairs of operators $\hat A$ and $\hat B$.   We can form the product $\hat A\hat B$.  We can take the
tensor product $\hat A \otimes \hat B$.  A third slightly more subtle example is $\hat A ? \hat B$ where $?$ stands
for an unknown operator. The first and third of these two examples correspond to a time-like situation whereas the
second case corresponds to a space-like situation (or at least an equal time situation).  Each of these cases is
treated on a different footing in QT. In GR we initially treat space and time on an equal footing. Thus, we
introduce four space-time coordinates $x^\mu$ (with $\mu = 0, 1, 2, 3$) giving rise to the intervals $\delta
x^\mu$.  We do this without saying which of these intervals (or which linear combinations of them) are time-like.
We then solve Einstein's field equations and obtain a metric $g_{\mu\nu}$. From the metric (which has a Lorentzian
signature) we can identify which linear combinations pertain to time-like intervals.   But, at least in principle,
we do this after we have solved the field equations.   Thus, by analogy, if we are to treat space and time on an
equal footing in QG as we do in GR then we would also want to put those objects in the theory of QG which
correspond to the three types of product in QT mentioned above on an equal footing. This should already be an issue
in special relativistic quantum theory though since the causal structure is fixed in advance it is not essential we
attend to it. But in QG it is likely to be quite essential.  In fact, in QG the issue is likely to be even more
serious than it is in GR since it may be, as mentioned above, that even after the equations have been solved, we
are unable to identify what intervals are space-like and what are time-like. In order to address this issue we will
define a new type of product which unifies all these types of products in quantum theory (and their counterparts in
more general probabilistic frameworks) putting them on an equal footing.

We should ultimately be interested in experiments to test a theory of QG.  Before we get to real experiments it is
interesting to consider gedankenexperiments which illuminate the conceptual structure of a theory.  As we noted in
the introduction, a theory of QG would be necessary in a situation in which we could not neglect the particular
effects of both GR and QT.   The type of gedankenexperiment in which this is going to be the case will be one where
we simultaneously have dynamic causal structure and quantum superposition.   Such a situation occurs when we look
for quantum interference of a massive object which goes into a superposition of being in two places at once.
Gedankenexperiments of this type have been discussed by Penrose \cite{Penrose} and there has been some effort to
design a realizable version of this type of experiment \cite{expmtPenrose}.

\section{Setting the scene}\label{meatstarts}

We repeat our assertion from Sec.\ \ref{Datumism}: A physical theory, whatever else it does, must correlate
recorded data.   Data is a record of actions and observations taken during an experiment.  We will assume that this
data is recorded onto cards.   Each card will record a small amount of {\it proximate} data. We will illustrate
what we mean by this soon with examples.  Thus the cards represent something analogous to Einsteinian events. One
piece of data recorded on any given card will be something we will regard as representing or being analogous to
space-time location.   Of course, it is not necessary to record the data onto cards.  It could be recorded in a
computer's memory, or by any other means.  But this story with the cards will help us in setting up the
mathematical framework we seek.

\begin{figure*}[t]
\resizebox{\textwidth}{!} {\includegraphics{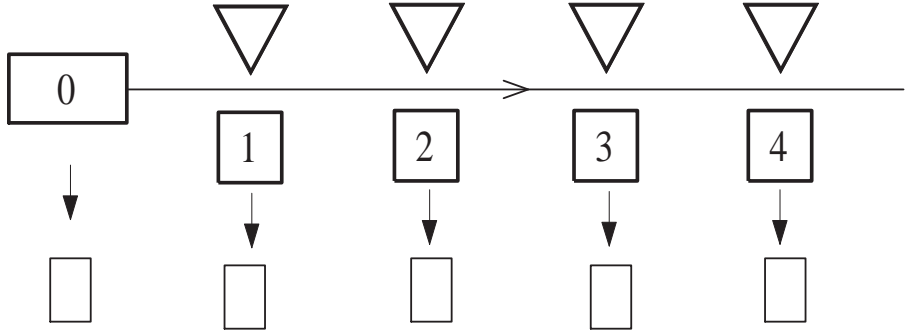}} \caption{\small A source of electrons is followed by four
Stern-Gerlach apparatuses shown schematically here. Data is recorded onto a card at each apparatus.}\label{SG}
\end{figure*}

\begin{figure*}[t]
\resizebox{\textwidth}{!} {\includegraphics{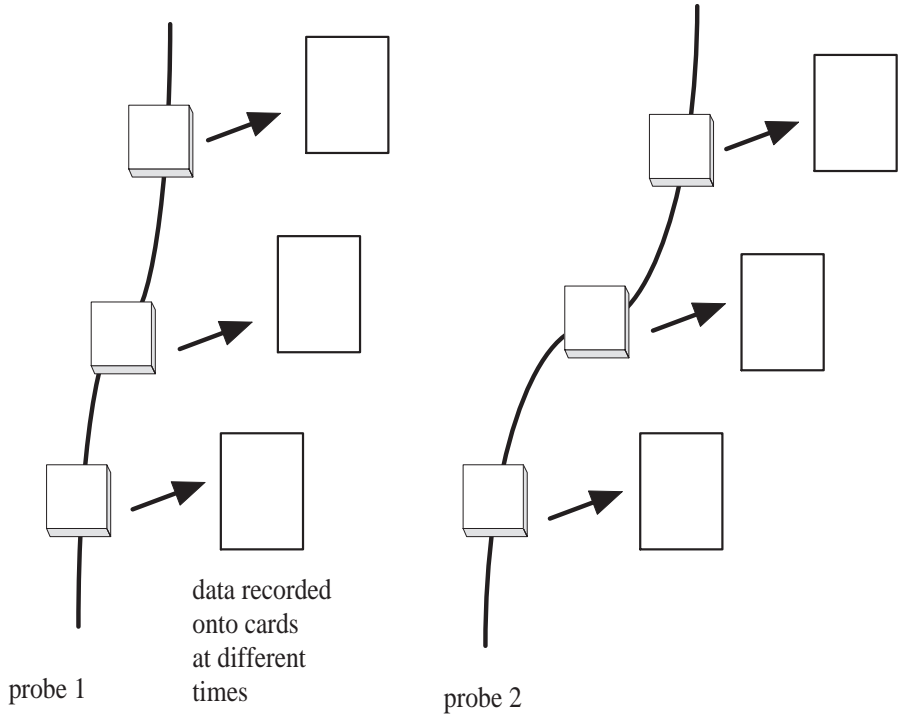}} \caption{\small We have a number of probes (two shown
here) drifting in space.  Data is collected onto a card at each probe at each tick of the probe's
clock.}\label{probes}
\end{figure*}

We will consider examples where the data recorded on each card is of the form $(x, a, s)$.  The first piece of
data, $x$, is an observation and represents location.   For example, it could be the space-time position read off
some real physical reference frame such as a GPS system \cite{GPS} or a reference mollusc \cite{popEinstein}.  Or
it might be some other data that we are simply going to regard as representing location.   The second piece of
data, $a$, represents some actions. For example it might correspond to the configuration of some knobs we have
freedom in setting.   The third piece of data, $s$, represents some further local observations.  Here are two
examples of this type
\begin{enumerate}
\item  Imagine we have a sequence of four Stern-Gerlach apparatuses (shown schematically in Fig.\ \ref{SG})
labelled $x=1, 2, 3, 4$ preceded by a source of electrons labelled $x=0$.  We can set each Stern-Gerlach apparatus
to measure spin along the direction $a$.  Then we record the outcome $s=\pm 1/2$.  Thus we get a card from each
Stern-Gerlach apparatus with $(x, a, s)$ written on it.  We would also want to extract a card (or maybe a set of
cards) from the apparatus which prepares the electrons.  For example we could write $(0, a, s)$ where $a=$ \lq\lq
source appropriately constructed", and s=\lq\lq observations consistent with source's proper functioning seen".
From each run of the experiment we would collect a stack of five cards.  We could vary the settings $a$ on the
Stern-Gerlach apparatuses.   To extract probabilities we would want to run the experiment many times.
\item  Imagine we have a number of probes labelled $n=1, 2, \dots N$ drifting in space (see Fig.\ \ref{probes}).
Each probe has a clock which ticks out times $t_n= 1, 2, \dots, T$.  Each probe has knobs on it which correspond to
the settings of some measurement apparatuses on the probe. We let the different configurations of these knobs be
labelled by $a$. Further, each probe has some meters which record the outcomes of those experiments. We let $s$
label these outcomes. On each card we can record the data $(x=(n, t_n, t^n_m), a, s)$ where $t^n_m$ is the retarded
time of probe $m$ as seen at probe $n$.  We may want to put more information into $x$, such as the observed
relative orientations.  And we may want to reduce the amount of information in $x$, for example we could only
record the retarded times of three specific probes (providing something like a GPS system).  We take one card from
each probe for each tick of the probe clock. At the end of each run of the experiment we will have a stack of $NT$
cards. The clocks can be reset and the experiment repeated many times to obtain probabilities.
\end{enumerate}
In these cases the number of cards in the stack is the same from one run to the next but this need not be the case.

We introduce the important notion of a {\it procedure}, $F$, which tells the experimentalist what actions to
implement - that is how to set the knobs.   In our examples above (where we record data $(x, a, s)$ onto the cards)
we allow the choice of $a$ to depend on $x$ by some function $a=F(x)$. After a procedure has been implemented we
will end up with a stack, $X$, of cards. This stack contains all the data for the experiment.

We may have more complicated ways of recording data onto cards.  A useful example is where we record $(x, q, a, s)$
onto the cards where $q$ represents some observations we do not want to regard as part of $x$ which we make
immediately before implementing the action $a$.  In this case we can have $a=F(x,q)$.

We may wish to condition $F$ at $x$ on some data collected \lq\lq previously" at $x'$.  For example we might want
to have $a=F(x, s')$ where $s'$ is data recorded at $x'$.  However, since we do not want to assume we have fixed
causal structure, it is a matter of the physical dynamics as to whether data recorded at $x'$ will be available at
$x$. Therefore it makes no sense to allow this functional dependence. Rather, any such dependence must be
implemented physically. For example, $q$ could be equal to the retarded value of $s'$ seen at $x$. Then we can
write $a=F(x,q)$.

It is possible that some cards in the stack will be repeated. This could happen for example if the clocks in the
second example above were faulty and sometimes ticked out the same value twice.  To get round this we replace
repeated cards by a new card having the same data plus the multiplicity appended to $s$.

\section{Thinking inside the box}

\begin{figure*}[t]
\resizebox{\textwidth}{!} {\includegraphics{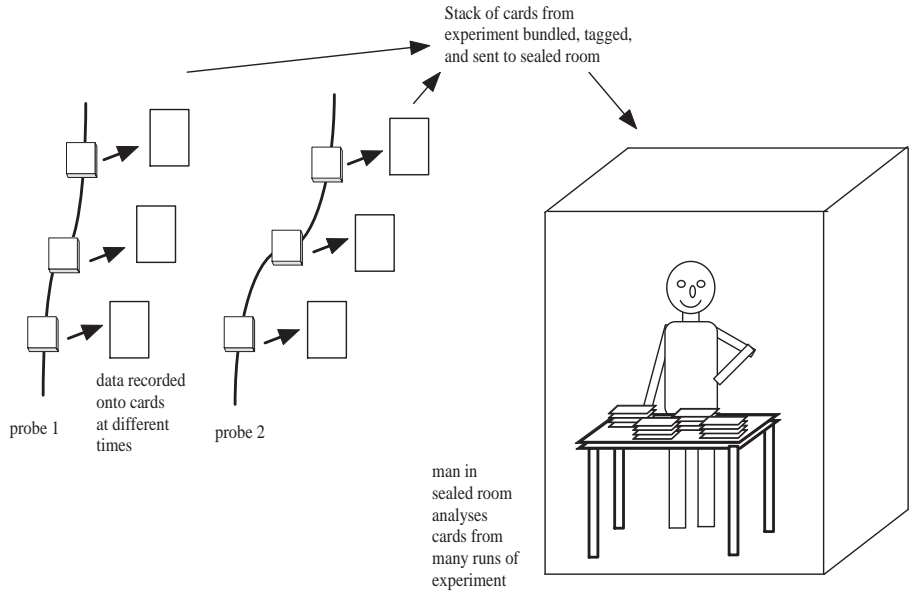}} \caption{\small Data is collected on cards from an
experiment, bundled and tagged with a description of the procedure followed. This tagged bundle is sent to a man in
a sealed room. This is repeated many times for each procedure.}\label{sealedrm}
\end{figure*}

After each run of the experiment we will have a {\it stack} of cards which we denote by $X$.  We can bundle these
together and attach a tag to this bundle specifying $F$.   This can be repeated many times for each possible $F$.
We imagine that all these tagged bundles are sent to a man in a sealed room who analyzes them (see Fig.\
\ref{sealedrm}).  The man cannot look outside the sealed room for extra clues.  Hence, all concepts must be defined
in terms of the cards themselves.  The point of this story is that it enforces a particular kind of honesty.  The
man is not able to introduce what Einstein called \lq\lq factitious causes" \cite{Einstein} such as an unobservable
global inertial reference frame since he is forced to work with the cards.

The order of the cards in any particular stack does not, in itself, constitute recorded data and is of no
significance.  Likewise the order the bundled stacks arrive in the sealed room is also of no significance.  Thus,
in his analysis of the cards, the man in the sealed room should not use these orderings.  We could imagine that
each stack is shuffled before being bundled and the bundles are then shuffled before being sent to the sealed room.
Of course, it is significant which bundle a particular card belongs to and so we should not allow cards from
different bundles to get mixed up.

To aid our analysis we begin with a few simple definitions.
\begin{quote}
{\bf The stack}, $X$, is the set of cards collected in one run of an experiment.
\end{quote}
\begin{quote}
{\bf  The full pack}, $V$ is the set of all logically possible cards when all possible procedures are taken into
account.  It is possible that some cards never actually occur in any stack because of the nature of the physical
theory but we include them in $V$ anyway.
\end{quote}
\begin{quote}
{\bf The procedure} will be specified by that set of cards $F$ which is the subset of all cards in $V$ which are
consistent with the procedure $F$.   For example, if the data on each card is of the form $(x, a, s)$ then the set
$F$ is all cards of the form $(x, F(x), s)$.   We deliberately use the same symbol, $F$, to denote the abstract
procedure $F$, the function $F(x)$, and the set $F$ since it will be clear from context which meaning is intended.
\end{quote}
We have
\begin{equation}
X\subseteq F \subseteq V
\end{equation}
We note that these definitions are in terms of the cards as required.

As described we imagine repeating the experiment many times.  In Sec.\ \ref{universalc} we will suggest an approach
that does not involve repeating the experiment as suggested here.

\section{Regions}

We continue to provide definitions in terms of the cards.   We define the notion of a {\it region}.  The region
$R_{\cal O}$ is specified by the set of cards from $V$ having $x\in{\cal O}$.   We define $R_x$ to be an {\it
elementary region} consisting only of the cards having $x$ on them.  Regions can be regarded as the seat of
actions.  In a region we have an independent choice of which action to implement.  This captures the notion of
space-time regions as places where we have local choices.

When we have a particular run of the experiment we end up with a stack $X$ of data.  We can allocate this data to
the appropriate regions.  Then we have a picture of what happened as laid out in a kind of space-time.  For a
region $R_1$ (which we take to be shorthand for $R_{{\cal O}_1}$) we define the stack in $R_1$ to be
\begin{equation}
X_{R_1}=X\cap R_1
\end{equation}
These are the cards from the stack, $X$,that belong to region $R_1$.

We define the procedure in $R_1$ to be
\begin{equation}
F_{R_1}=F\cap R_1
\end{equation}
these are the cards from the set $F$ which belong to $R_1$.  Clearly
\begin{equation}
X_{R_1} \subseteq F_{R_1}\subseteq R_1
\end{equation}
Given $(X_{R_1}, F_{R_1})$ we know \lq\lq what was done" ($F_{R_1}$) and \lq\lq what was seen" ($X_{R_1}$) in
region $R_1$.

\section{Statement of objective}\label{statement}

We seek to find a formalism to calculate conditional probabilities of the form
\begin{equation}
{\rm prob}(X_{R_1}| X_{R_2}, F_{R_1}, F_{R_2})
\end{equation}
when these probabilities are well defined without imposing any particular causal structure in advance.   Of course,
any particular theory we might cast in terms of this formalism to is likely to have some sort of causal structure
built in and this will be evident in the particular form the mathematical objects in the formalism end up taking.

If the above probability is given a frequency interpretation it is equal to
\begin{equation}
\frac{N(X_{R_1}, X_{R_2}, F_{R_1}, F_{R_2})}{N(X_{R_2}, F_{R_1}, F_{R_2})}
\end{equation}
in the limit as the denominator becomes large.  Here $N(\cdot)$ is the number of stacks satisfying the given
condition.

This probability may not be well defined if there is insufficient conditioning.  To see this consider the example
with four Stern-Gerlach apparatuses illustrated in Fig.\ \ref{SG}. Let $R_1$ consist of all the cards associated
with the fourth Stern-Gerlach apparatus (having $x=4$) and let $R_2$ consist of all the cards associated with the
second apparatus (having $x=2$).  Then we cannot expect this probability to be well defined since we do not know
what angle the third Stern-Gerlach apparatus has been set at.  In such cases we do not require the formalism to
predict any value for the probability.

To aid our considerations we will restrict our attention to a region $R$ for which all probabilities
\begin{equation}
{\rm prob}(X_{R}| F_{R}, C)
\end{equation}
are well defined where $C$ is some condition on the cards outside $R$. We will call such a region, $R$, a {\it
predictively well defined region} with respect to the conditioning $C$.   This region can be very big (consisting
of a substantial fraction of the cards from $V$). We will assume that we are only interested in probabilities
inside this region and, since it is always implicit, we will drop the $C$ writing
\begin{equation}\label{probXF}
{\rm prob}(X_{R}| F_{R})
\end{equation}
An example of a predictively well defined region might be provided by the data coming out of a quantum optical
laboratory.  We would have to set up the laboratory, the optical table, the lasers, the various optical elements
and the computers to record data. Having set this up we would want to keep the doors of the laboratory shut, the
shutters on the windows down, and condition on many other actions and observations to isolate the experiment from
stray light. All this data could go into $C$.  In setting up the formalism we will assume that $C$ is always
satisfied and that it is the purpose of a physical theory to correlate what goes on inside $R$.  Having set up the
formalism we will discuss ways to avoid having to have a predictively well defined region and having to
conditionalize on $C$ (see Sec.\ \ref{opencausaloid} and Sec.\ \ref{noboundary}).

\section{States}

\begin{figure*}[t]
\resizebox{\textwidth}{!} {\includegraphics{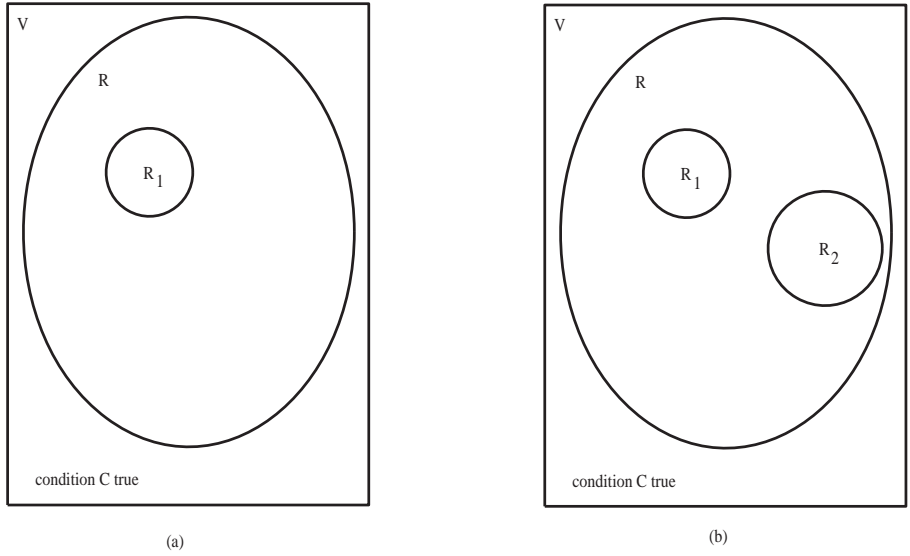}} \caption{\small (a) Region $R_1$ inside the predictably well
defined region $R$.  (b) Regions $R_1$ and $R_2$ inside predictably well defined region $R$.}\label{R1R2}
\end{figure*}

Now consider region $R_1$ inside the predictively well defined region $R$ (see Fig.\ \ref{R1R2}(a)).   Then
(\ref{probXF}) can be written
\begin{equation}\label{split}
p={\rm prob}(X_{R_1}\cup X_{R-R_1}| F_{R_1}\cup F_{R-R_1})
\end{equation}
We will regard $(X_{R-R_1}, F_{R-R_1})$ which happens in $R-R_1$ as a {\it generalized preparation} for region
$R_1$. Associated with this generalized preparation for $R_1$ is a state which we will define shortly.  We will
regard $(X_{R_1}, F_{R_1})$ which happens in $R_1$ as a {\it measurement}.   We will label measurements in $R_1$
with $\alpha_1$. Then we can write
\begin{equation}\label{splitalpha}
p_{\alpha_1}={\rm prob}(X^{\alpha_1}_{R_1}\cup X_{R-R_1}| F^{\alpha_1}_{R_1}\cup F_{R-R_1})
\end{equation}
We define
\begin{quote}
{\bf The state} for $R_1$ associated with a generalized preparation in $R-R_1$ is defined to be that thing
represented by any mathematical object which can be used to predict $p_{\alpha_1}$ for all ${\alpha_1}$.
\end{quote}
Note that quite deliberately we define this for the joint probabilities in (\ref{splitalpha}) rather than the
conditional probabilities ${\rm prob}(X^{\alpha_1}_{R_1}| X_{R-R_1}, F^{\alpha_1}_{R_1}\cup F_{R-R_1})$ even though
the latter may seem more natural.   The reason for this is that introducing conditional probabilities requires
normalization by Bayes formula which introduces nonlinearities.  It turns out that these nonlinearities would
represent an insurmountable problem in the case where we have dynamic causal structure and so it is better to work
with the joint probabilities. As we will see, we can use Bayes formula in the final step when calculating
conditional probabilities so there is no problem.

Given the above definition we could write the state as
\begin{equation}\label{alphaeqn}
{\bf P}(R_1)= \left( \begin{array}{c} \vdots \\ p_{\alpha_1} \\ \vdots \end{array} \right)
\end{equation}
We can then write
\begin{equation}\label{RdotP}
p_{\alpha_1}= {\bf R}_{\alpha_1}(R_1)\cdot {\bf P}(R_1)
\end{equation}
where the vector ${\bf R}_{\alpha_1}(R_1)$ has a 1 in position ${\alpha_1}$ and 0's in all other positions.

Now, we would expect a physical theory to have some structure such that it is not necessary to list all
probabilities as in ${\bf P}(R_1)$ but rather only some subset of them. Thus, we pick out a set of {\it fiducial}
measurements $(X^{k_1}_{R_1}, F^{k_1}_{R_1})$, where $k_1\in\Omega_1$, {\it such that we can write a general
probability by means of a linear formula}
\begin{equation}\label{rdotp}
p_{\alpha_1}= {\bf r}_{\alpha_1}(R_1)\cdot {\bf p}(R_1)
\end{equation}
where
\begin{equation}
{\bf p}(R_1)= \left( \begin{array}{c} \vdots \\ p_{k_1} \\ \vdots \end{array} \right) {\rm ~~where~~}
k_1\in\Omega_1
\end{equation}
now represents the state with
\begin{equation}
p_{k_1} = {\rm prob}(X^{k_1}_{R_1}\cup X_{R-R_1}| F^{k_1}_{R_1}\cup F_{R-R_1})
\end{equation}
and where $K_1=|\Omega_1|$ is taken to be the minimum number of probabilities in such a list.  It is clear we can
always do this since, as a last resort, we have (\ref{RdotP}).   Omega sets such as $\Omega_1$ will play a big role
in this paper.  The are not, in general, unique but we can always pick one and stick with it.

The vector ${\bf r}_{\alpha_1}(R_1)$ is associated with the measurement $(X^{\alpha_1}_{R_1}, F^{\alpha_1}_{R_1})$
in $R_1$. The fiducial measurements are represented by
\begin{equation}\label{fidmeas}
{\bf r}_{k_1} =\left( \begin{array}{c} 0 \\ 0 \\ \vdots \\ 1 \\ \vdots \\ 0 \end{array} \right) {\rm ~~for~all~~}
{k_1}\in\Omega_1
\end{equation}
where the 1 is in the ${k_1}$ position since this is the only way to ensure that ${\bf r}_{k_1}\cdot {\bf p} =
p_{k_1}$ as required.

We define $\Lambda_{\alpha_1}^{k_1}$ by
\begin{equation}
{\bf r}_{\alpha_1}= \sum_{k_1\in\Omega_1} \Lambda_{\alpha_1}^{k_1} {\bf r}_{k_1}
\end{equation}
We will define further lambda matrices in the next section.  Like omega sets, they will play a central role in this
work.  They give a quantitative handle on the amount of compression the physical theory provides.  Given
(\ref{fidmeas}) it is clear that here the lambda matrices are just the components of the vector ${\bf
r}_{\alpha_1}$. That is
\begin{equation}\label{rlambda}
{\bf r}_{\alpha_1}|_{k_1}\equiv r_{k_1}^{\alpha_1} = \Lambda_{\alpha_1}^{k_1}
\end{equation}
We will sometimes drop the $\alpha$'s writing ${\bf r}$ with components $r_{k_1}$.   The $\alpha$'s are then
understood to be implicit.   Note it also follows from the definition of the $\Lambda$ matrix that
\begin{equation}\label{lambdadelta}
\Lambda_{k'_1}^{k_1} = \delta_{k'_1}^{k_1} {\rm ~~~for~~~} k'_1, k_1 \in \Omega_1
\end{equation}
where $\delta_{k'_1}^{k_1}$ equals 1 if the subscript is equal to the superscript and 0 otherwise.

\section{Composite Regions}

Now consider two disjoint regions $R_1$ and $R_2$ in $R$ (see Fig.\ \ref{R1R2}(b)).  We have
\begin{eqnarray}
\lefteqn{ {\rm prob}(X_{R_1}\cup X_{R_2}\cup X_{R-R_1-R_2}   |  F_{R_1}\cup F_{R_2}\cup F_{R-R_1-R_2})  }
\qquad\qquad\qquad\qquad\qquad\qquad \nonumber\\
&=&{\bf r}(R_1)\cdot {\bf p}(R_1)  \nonumber \\
&=&\sum_{k_1} r_{k_1}(R_1) p_{k_1}(R_1)   \nonumber \\
&=&\sum_{k_1} r_{k_1}(R_1) {\bf r}(R_2)\cdot {\bf p}_{k_1}(R_2) \nonumber \\
&=& \sum_{k_1k_2} r_{k_1} r_{k_2} p_{k_1k_2} \label{mainproof}
\end{eqnarray}
where ${\bf p}_{k_1}(R_2)$ is the state in $R_2$ given the generalized preparation $(X^{k_1}_{R_1}\cup
X_{R-R_1-R_2}, F^{k_1}_{R_1}\cup F_{R-R_1-R_2})$ in region $R-R_2$ and where
\begin{equation}
p_{k_1k_2} = {\rm prob}(X^{k_1}_{R_1}\cup X^{k_2}_{R_2} \cup X_{R-R_1-R_2} | F^{k_1}_{R_1}\cup F^{k_2}_{R_2} \cup
F_{R-R_1-R_2})
\end{equation}
This means that we can write the probability for any measurement for the composite region $R_1\cup R_2$ in terms of
a linear sum of joint probabilities $p_{k_1k_2}$ with $k_1k_2\in \Omega_1\times\Omega_2$.  It may even be the case
that do not need all of these probabilities. There may be some further compression (though still maintaining that
we have a linear sum). Hence we have the result that a fiducial set of measurements for the composite region
$R_1\cup R_2$ is given by
\begin{equation}
(X^{k_1}_{R_1}\cup X^{k_2}_{R_2}, F^{k_1}_{R_1}\cup F^{k_2}_{R_2}) ~~{\rm for}~~k_1k_2\in \Omega_{12}\subseteq
\Omega_1\times\Omega_2
\end{equation}
This result is the central to the whole approach in this paper.

If the behaviour in the two regions is not causally connected then we expect that
$\Omega_{12}=\Omega_1\times\Omega_2$.  On the other hand, if there is a strong causal connection then we can have
$|\Omega_{12}|=|\Omega_1|=|\Omega_2|$.  The relationships between these $\Omega$ sets gives us a combinatoric
handle on the causal structure.  We seek, however, a more quantitative handle.

To this end we define $\Lambda_{l_1l_2}^{k_1k_2}$ by
\begin{equation}\label{defnlambda12}
{\bf r}_{l_1l_2} = \sum_{k_1k_2\in\Omega_{12}} \Lambda_{l_1l_2}^{k_1k_2} {\bf r}_{k_1k_2} {\rm ~~for~~}
l_1l_2\in\Omega_1\times \Omega_2
\end{equation}
We will adopt the convention of labelling the elements of the post compression omega set with $k$'s and the
elements of the pre-compression product set with $l$'s as in this equation. As before, the fiducial measurements
are represented by
\begin{equation}\label{fidmeascomp}
{\bf r}_{k_1k_2} =\left( \begin{array}{c} 0 \\ 0 \\ \vdots \\ 1 \\ \vdots \\ 0 \end{array} \right) {\rm
~~for~all~~} {k_1k_2}\in\Omega_{12}
\end{equation}
where the 1 is in the ${k_1k_2}$ position.  It follows by taking the $k_1k_2$ component of (\ref{defnlambda12})
that
\begin{equation}\label{r12lambda12}
{\bf r}_{l_1l_2}|_{k_1k_2} \equiv r^{l_1l_2}_{k_1k_2} =  \Lambda_{l_1l_2}^{k_1k_2}
\end{equation}
This is similar to (\ref{rlambda}) above.

We can use lambda matrices to calculate an arbitrary ${\bf r}$ for the composite system.   To see this we start by
putting (\ref{rlambda}) in (\ref{mainproof}) and reinserting $\alpha$'s
\begin{eqnarray}
{\bf r}_{\alpha_1\alpha_2}\cdot {\bf p}& = & \sum_{l_1l_2\in\Omega_1\times\Omega_2} \Lambda_{\alpha_1}^{l_1}
\Lambda_{\alpha_2}^{l_2} p_{l_1l_2} \\
&=& \sum_{l_1l_2\in\Omega_1\times\Omega_2} \Lambda_{\alpha_1}^{l_1}
\Lambda_{\alpha_2}^{l_2} {\bf r}_{l_1l_2}\cdot {\bf p} \\
&=& \sum_{l_1l_2\in\Omega_1\times\Omega_2} \Lambda_{\alpha_1}^{l_1} \Lambda_{\alpha_2}^{l_2}
\sum_{k_1k_2\in\Omega_{12}} \Lambda_{l_1l_2}^{k_1k_2} {\bf r}_{k_1k_2}\cdot {\bf p}
\end{eqnarray}
Since this must be true for all ${\bf p}$ we have,
\begin{equation}
{\bf r}_{\alpha_1\alpha_2} = \sum_{k_1k_2\in\Omega_{12}} \left(\sum_{l_1l_2\in\Omega_1\times\Omega_2}
\Lambda_{\alpha_1}^{l_1} \Lambda_{\alpha_2}^{l_2}  \Lambda_{l_1l_2}^{k_1k_2}\right) {\bf r}_{k_1k_2}
\end{equation}
and hence, in view of (\ref{fidmeascomp}) above, the components of ${\bf r}_{\alpha_1\alpha_2}$ are
\begin{equation}\label{rlambda2}
{\bf r}_{\alpha_1\alpha_2}|_{k_1k_2} = \sum_{l_1l_2\in\Omega_1\times\Omega_2} \Lambda_{\alpha_1}^{l_1}
\Lambda_{\alpha_2}^{l_2}  \Lambda_{l_1l_2}^{k_1k_2}
\end{equation}
This is consistent with (\ref{r12lambda12}) above given (\ref{lambdadelta}).

We can generalize this for $N$ regions.  We define lambda matrices for multi-region composites by
\begin{equation}
{\bf r}_{l_1\dots l_N} = \sum_{k_1\dots k_N\in\Omega_{12}} \Lambda_{l_1\dots l_N}^{k_1\dots k_2} {\bf r}_{k_1\dots
k_2} {\rm ~~for~~} l_1\dots l_2\in\Omega_1\times\dots \times \Omega_N
\end{equation}
And then it is easy to show that
\begin{equation}\label{rmultilambda}
{\bf r}_{\alpha_1\cdots \alpha_N}|_{k_1\cdots k_N} = \sum_{l_1\cdots l_N\in\Omega_1\times\cdots \times\Omega_N}
\Lambda_{\alpha_1}^{l_1} \Lambda_{\alpha_2}^{l_2}\cdots \Lambda_{\alpha_N}^{l_N}  \Lambda_{l_1\cdots
l_N}^{k_1\cdots k_2}
\end{equation}
Hence, given the lambda matrices we have a way of calculating the components of ${\bf r}$ vectors for one region
(\ref{rlambda}), two regions (\ref{rlambda2}), and multi-regions (\ref{rmultilambda}).

\section{The Causaloid}\label{causaloid}

In the previous section we discussed composite regions made from regions $R_1$, $R_2$, \dots.  The smallest
component regions are the elementary regions $R_x$.  Any region, $R_{\cal O}$, is composed of elementary regions. A
general measurement in this region is labelled with $\alpha_{\cal O}$.  But since a general measurement decomposes
into local measurements at each component elementary region we can write
\begin{equation}
\alpha_{\cal O} = \alpha_{x}\alpha_{x'}\cdots \alpha_{x''}  {\rm ~~where ~~} {\cal O}= \{ x,x', \cdots x''\}
\end{equation}
For each of these elementary regions we will have a local lambda matrix $\Lambda_{\alpha_x}^{k_x}(x, \Omega_x)$
with $k_x\in\Omega_x$.  We include the argument $x$ for clarity and the argument $\Omega_x$ since the choice of
omega set is not, in general, unique.   For the region $R_{\cal O}$ we have lambda matrix
\begin{equation}\label{lambdaO}
\Lambda_{l_{\cal O}}^{k_{\cal O}}({\cal O}, \Omega_{\cal O})
\end{equation}
where
\begin{equation}\label{omegaprodxs}
l_{\cal O} \equiv l_{x} l_{x'}\cdots l_{x''}\in\Omega_{x}\times\Omega_{x'}\times \cdots\times \Omega_{x''} {\rm
~~where~~}  {\cal O}= \{ x,x', \cdots x''\}
\end{equation}
and
\begin{equation}\label{omegakxs}
k_{\cal O} \equiv k_{x} k_{x'}\cdots k_{x''}\in\Omega_{\cal O} {\rm ~~where~~} {\cal O}= \{ x,x', \cdots x''\}
\end{equation}
We will sometimes have reason to consider a lambda matrix as in (\ref{lambdaO}) but where ${\cal O}=\{ x \}$. That
is $\Lambda_{l_x}^{k_x}$.  By convention $l$'s are in the product set and $k$'s are in the new omega set.  But in
this case there is only one omega set, namely $\Omega_x$ to take a product over.  Thus, we have
\begin{equation}\label{elementarylambda}
\Lambda_{l_x}^{k_x} = \delta_{l_x}^{k_x} {\rm ~~~ with ~~~} l_x, k_x \in \Omega_x
\end{equation}
It is worth noting that if we know $\Lambda_{l_{\cal O}}^{k_{\cal O}}({\cal O}, \Omega_{\cal O})$ for one omega
set then, since the lambda matrix contains all relevant linear dependencies, we can calculate (i) all other omega
sets and (ii) the lambda matrix for any other omega set for the given region (we will spell out the method for
doing this in Sec.\ \ref{transforming}). Hence, it is enough to know the lambda matrix for one omega set for each
region. The lambda matrices can be used to calculate an arbitrary measurement vector using the results of the
previous sections applied to elementary regions.  From (\ref{rlambda}) and (\ref{rmultilambda})
\begin{equation}\label{rmultilambdax}
{\bf r}_{\alpha_x}|_{k_x} = \Lambda_{\alpha_x}^{k_x}  ~~~~~~~~ {\bf r}_{\alpha_{\cal O}}|_{k_{\cal O}} =
\sum_{l_{\cal O}} \Lambda_{\alpha_x}^{l_x} \Lambda_{\alpha_{x'}}^{l_{x'}}\cdots \Lambda_{\alpha_{x''}}^{l_{x''}}
\Lambda_{l_{\cal O}}^{k_{\cal O}}
\end{equation}

We now come to the central mathematical object in the approach to be taken in this paper.
\begin{quote}
{\bf The causaloid } for a predictively well defined region $R$ made up of  elementary regions $R_x$ is defined to
be that thing represented by any mathematical object which can be used to obtain ${\bf r}_{\alpha_{\cal O}}(R_{\cal
O})$ for all measurements $\alpha_{\cal O}$ in region $R_{\cal O}$ for all $R_{\cal O}\subseteq R$.
\end{quote}
In view of the above results, one mathematical object which specifies the causaloid is
\begin{equation}\label{causaloideqn}
\left[
\begin{array}{ll} \Lambda_{\alpha_x}^{k_x}(x, \Omega_x) &: {\rm for~one} ~~\Omega_x ~~{\rm for~each} ~~ R_x \\
\Lambda^{k_{\cal O}}_{l_{\cal O}}({\cal O}, \Omega_{\cal O})&: {\rm for~one} ~~\Omega_{\cal O} ~\text{for each
non-elementary} ~ R_{\cal O} \subseteq R
\end{array}
\right]
\end{equation}
This lists all $\Lambda$ matrices.  However, we might expect any given physical theory to have some structure such
that some $\Lambda$ matrices can be calculated from others.  If this is the case then we might expect that we can
take some subset of the $\Lambda$ matrices, labelled by $j$ and write
\begin{equation}
{\bf \Lambda}= [ \Lambda(j): j=1~{\rm to}~ J| {\rm RULES}]
\end{equation}
where RULES are rules for deducing a general $\Lambda$ from the given $\Lambda$'s.  We will see that we can
achieve quite considerable compression of this nature in the cases of CProbT and QT.

\section{The causaloid product}

As we noted in Sec.\ \ref{QGsec}, there are three basic ways of putting two operators together in quantum theory:
$\hat A \hat B$, $\hat A ? \hat B$, and $\hat A\otimes \hat B$.  We noted there that it would be desirable to
treat these on an equal footing.  To this end we now define the {\it causaloid product} for our framework. At this
stage we are working in a general framework. However, we will see later that this unifies all these types of
product for quantum theory.  Let ${\bf r}_{\alpha_1}$ be a measurement vector in $R_1$ (corresponding to ${\cal
O}_1$) and let ${\bf r}_{\alpha_2}$ be a measurement vector in $R_2$ (corresponding to ${\cal O}_2$) and let the
regions $R_1$ and $R_2$ be non-overlapping.  We define the causaloid product $\otimes^{\Lambda}$ by
\begin{equation}\label{causaloidprod}
{\bf r}_{\alpha_1\alpha_2}= {\bf r}_{\alpha_1}\otimes ^{\Lambda} {\bf r}_{\alpha_2}
\end{equation}
Strictly we should write
\begin{equation}
({\bf r}_{\alpha_1\alpha_2}, {\cal O}_1\cup{\cal O}_2) = ({\bf r}_{\alpha_1}, {\cal O}_1) \otimes ^{\Lambda} ({\bf
r}_{\alpha_2}, {\cal O}_2)
\end{equation}
since the causaloid product needs to know which region it is addressing but for brevity we will stick with
(\ref{causaloidprod}) the regions being implicit in the labels $\alpha_1$ and $\alpha_2$.  The components of the
LHS are obtained from the components of the vectors on the RHS by applying (\ref{rlambda}) and (\ref{rlambda2})
\begin{equation}\label{genproduct}
{\bf r}_{\alpha_1\alpha_2}|_{k_1k_2} = \sum_{l_1l_2\in\Omega_1\times\Omega_2} ({\bf r}_{\alpha_1}|_{l_1}) ({\bf
r}_{\alpha_2}|_{l_2}) \Lambda_{l_1l_2}^{k_1k_2}
\end{equation}
Now, the lambda matrix $\Lambda_{l_1l_2}^{k_1k_2}$ is given by the causaloid.  To see this we can reinsert the
${\cal O}$'s writing it as $\Lambda_{l_{{\cal O}_1}l_{{\cal O}_2}}^{k_{{\cal O}_1}k_{{\cal O}_2}}$ which, in view
of (\ref{omegaprodxs}, \ref{omegakxs}), is the same as $\Lambda_{l_{{\cal O}_1\cup {\cal O}_2}}^{k_{{\cal
O}_1\cup{\cal O}_2}}$. Now, in fact the causaloid gives this lambda matrix for the $l$'s in the product set over
all elementary product sets $\Omega_x$'s (as in (\ref{omegaprodxs})) whereas we only require $l$'s in the subset
$\Omega_{{\cal O}_1}\times\Omega_{{\cal O}_2}$. We use only those components of $\Lambda_{l_{{\cal O}_1\cup {\cal
O}_2}}^{k_{{\cal O}_1\cup{\cal O}_2}}$ we need to take the causaloid product.

We wish to discuss two special cases.
\begin{enumerate}
\item {\it Omega sets multiply.} $\Omega_{12}=\Omega_1\times\Omega_2$ so that
$|\Omega_{12}|=|\Omega_1||\Omega_2|$.
\item {\it Omega sets do not multiply.} $\Omega_{12}\subset\Omega_1\times\Omega_2$ so that $|\Omega_{12}|< |\Omega_1||\Omega_2|$.
\end{enumerate}
First note that it follows immediately from the definition of the lambda matrix in (\ref{defnlambda12}) that
\begin{equation}
\Lambda_{l_1l_2}^{k_1k_2} = \delta_{l_1l_2}^{k_1k_2} {\rm ~~for~~} l_1l_2, k_1k_2 \in \Omega_{12}
\end{equation}
where $\delta_{l_1l_2}^{k_1k_2}$ equals 1 if the subscripts and superscripts are equal and 0 otherwise.  Hence, we
can write (\ref{genproduct}) as
\begin{equation}\label{genproducte}
{\bf r}_{\alpha_1\alpha_2}|_{k_1k_2} =  ({\bf r}_{\alpha_1}|_{k_1}) ({\bf r}_{\alpha_2}|_{k_2}) +
\sum_{l_1l_2\in\Omega_1\times\Omega_2-\Omega_{12}} ({\bf r}_{\alpha_1}|_{l_1}) ({\bf r}_{\alpha_2}|_{l_2})
\Lambda_{l_1l_2}^{k_1k_2}
\end{equation}
It follows that
\begin{equation}\label{ifomegamultiply}
{\rm if}~~~\Omega_{12}=\Omega_1\times\Omega_2~~~{\rm then}~~~ {\bf r}_{\alpha_1\alpha_2} = {\bf
r}_{\alpha_1}\otimes{\bf r}_{\alpha_2}
\end{equation}
where $\otimes$ denotes the ordinary tensor product.  Hence we see that the ordinary tensor product is a special
case of the causaloid product when the omega sets multiply.  We will see that, in quantum theory, typically omega
sets will multiply.  This corresponds to the products $\hat A ? \hat B$ and $\hat A\otimes \hat B$ from quantum
theory which have the property that the total number of real parameters after taking the product is equal to the
product of the number from each operator (so we have $|\Omega_{12}|=|\Omega_1||\Omega_2|$).  Omega sets do not
multiply when we have {\it strong causal dependence} so that what happens in one region depends, at least
partially, on what is done in the other region in a way that cannot be altered by what is done in the rest of $R$.
In quantum theory we see this when we take the product $\hat A\hat B$.  Then the total number of real parameters in
the product is equal to the number in $\hat A$ and  $\hat B$ separately (this is basically
$|\Omega_{12}|=|\Omega_1|=|\Omega_2|$). Typically strong causal dependence indicates that two regions are
sufficiently \lq\lq close" that what is done outside of these regions cannot interfere with the causal dependence.
We will say that the two regions are {\it causally adjacent} in these cases.

\section{Using the causaloid to make predictions}\label{cpredictions}

The causaloid is so named because it is an object which gives us a quantitative handle on the causal structure as
was seen in the previous section.    What is surprising is that, given the causaloid, we can calculate any
probability pertaining to the predictively well defined region $R$ so long as that probability is well defined. To
see this note that if we have disjoint regions $R_1$ and $R_2$ we can write
\begin{equation}\label{bayes}
{\rm prob}(X_{R_1}|X_{R_2}, F_{R_1}, F_{R_2}) = \frac{ {\bf r}_{(X_{R_1},F_{R_1})}\otimes^{\Lambda} {\bf
r}_{(X_{R_2}, F_{R_2})} \cdot {\bf p}({R_1\cup R_2}) } { \sum_{ Y_{R_1}}{\bf
r}_{(Y_{R_1},F_{R_1})}\otimes^{\Lambda} {\bf r}_{(X_{R_2}, F_{R_2})} \cdot {\bf p}({R_1\cup R_2}) }
\end{equation}
(this is basically Bayes formula).  The sum in the denominator is over all possible stacks $Y_{R_1}$ in $R_1$
consistent with $F_{R_1}$, that is all $Y_{R_1}\subseteq F_{R_1}$. All probabilities pertaining to region $R$ are
of this form. Now, if this probability is well defined then it does not depend on what is outside $R_1\cup R_2$.
Hence, it does not depend on the state ${\bf p}({R_1\cup R_2})$. The space of possible states spans the full
dimensionality of the vector space by definition since we have a minimal fiducial set of measurements specifying
the components of the state. Hence
\begin{quote}
The probability
\[{\rm prob}(X_{R_1}|X_{R_2}, F_{R_1}, F_{R_2})   \]
is well defined if and only if
\[ {\bf v}\equiv {\bf r}_{(X_{R_1},F_{R_1})}\otimes^{\Lambda} {\bf r}_{(X_{R_2}, F_{R_2})}  \]
is parallel to
\[ {\bf u}\equiv \sum_{ Y_{R_1}\subseteq F_{R_1}}{\bf r}_{(Y_{R_1},F_{R_1})}\otimes^{\Lambda} {\bf r}_{(X_{R_2}, F_{R_2})}  \]
and this probability is given by
\begin{equation}\label{bayesparallel}
{\rm prob}(X_{R_1}|X_{R_2}, F_{R_1}, F_{R_2})  = \frac{|{\bf v}|}{|{\bf u}|}
\end{equation}
where $|{\bf a}|$ denotes the length of the vector ${\bf a}$.
\end{quote}
In fact, since the two vectors are parallel, we can simply take the ratio of any given component of the two vectors
as long as the denominator is nonzero.  We can write ${\bf v}= p {\bf u}$ where $p$ is the above probability.

One concern might be that it will be a rare state of affairs that these vectors are parallel and so the formalism
is rarely useful.  Since we have only set ourselves the task of calculating probabilities when they are well
defined we are not compelled to address this.  However, it turns out that the situation is not so bad.  In fact we
can always make $R_2$ big enough that we have well defined probabilities. To see this consider the extreme case
where $R_2=R-R_1$.   Then we have ${\bf p}({R_1\cup R_2})={\bf p}(R)$.   Now, we only have one preparation for the
predictively well defined region $R$, namely the condition $C$ being true on the cards outside $R$. Since we are
always taking this to be true we can only have one state.  This means that it must be specified by a single
component, that is ${\bf p}(R)=(p_1)$ where $(p_1)$ is a single component vector.  The number $p_1$ cancels out in
(\ref{bayes}) and so the probability is well defined.

If the two vectors are not exactly parallel the probability will not be well defined but it may be be approximately
well defined.  Indeed, we can use (\ref{bayes}) to place bounds on the probability.  Define ${\bf v}^\parallel$ and
${\bf v}^\perp$ as the components of ${\bf v}$ parallel and perpendicular to ${\bf u}$ respectively.   Then it is
easy to show that
\begin{equation}\label{pbounded}
\frac{|{\bf v}^\parallel|}{|{\bf u}|} - \frac{|{\bf v}^\perp|}{|{\bf v}|\cos\phi} \leq {\rm prob}(X_{R_1}|X_{R_2},
F_{R_1}, F_{R_2}) \leq \frac{|{\bf v}^\parallel|}{|{\bf u}|} + \frac{|{\bf v}^\perp|}{|{\bf v}|\cos\phi}
\end{equation}
where $\phi$ is the angle between ${\bf v}$ and ${\bf v}^\perp$ (we get these bounds using $|{\bf v}\cdot{\bf
p}|\leq |{\bf u}\cdot{\bf p}|$ and considering ${\bf p}$ parallel to ${\bf v}^\perp$).

\section{Physical theories and the causaloid formalism}\label{physicaltheories}

In the causaloid formalism
\begin{enumerate}
\item We have the causaloid, ${\bf \Lambda}$, which is theory specific.  The causaloid depends on the boundary conditions
$C$ outside $R$.  These might only be relevant when we are \lq\lq close" to the boundary (QT appears to be of this
nature).  In this case, modulo what happens at the boundary, we can say that the causaloid fully characterizes the
physical theory (at least its predictive component).
\item We have some basic equations which are theory non-specific.  These are (\ref{rmultilambdax})
for calculating a general ${\bf r}$ from the causaloid, (\ref{genproduct}) for forming the causaloid product,  and
Bayes formula in the form given in (\ref{bayes}) above (or we can use (\ref{bayesparallel}) given that appropriate
conditions are satisfied).
\end{enumerate}
Given the theory specific part and the theory independent equations we can make predictions.  This framework is
very general and we would expect any physical theory to fit into it (perhaps with some minor modifications
concerning the way the data is collected onto cards).   Hence we see that we have a potentially very powerful
formalism in which the theory specifics are all put into one object.   This is likely to be useful if we are hoping
to combine different physical theories.

\section{The open causaloid}\label{opencausaloid}

In typical situations we will have some elementary regions which can be regarded as being at the boundary.
Typically we might expect to have to use special mathematical techniques to deal with these.  However, if the
region $R$ is sufficiently big then we are most likely to be interested in probabilities which do not pertain to
the these boundary regions. For this reason we define the {\it open causaloid}.
\begin{quote}
{\bf The open causaloid } for a predictively well defined region $R$ made up of  elementary regions $R_x$ with
boundary elementary regions $R_{x_b}$ with $x_b\in{\cal O}_b$ is defined to be that thing represented by any
mathematical object which can be used to obtain ${\bf r}_{\alpha_{\cal O}}(R_{\cal O})$ for all measurements
$\alpha_{\cal O}$ in region $R_{\cal O}$ for all $R_{\cal O}\subseteq R-R_{{\cal O}_b}$.
\end{quote}
We can use the open causaloid to calculate all well defined probabilities excluding those which pertain to the
boundary. If we make the region $R$ sufficiently big then we can be sure that any regions like $R_1$ and $R_2$ of
interest (and for which we want to calculate conditional probabilities) do not overlap with the boundary regions.
In this case the open causaloid is as useful as the causaloid itself.  Indeed, given that we have already
restricted our attention to $R$ in defining the causaloid, it does not matter much if we restrict slightly further
to $R-R_{{\cal O}_b}$. In view of the remarks in the last section, the open causaloid is likely to be
characteristic of the physical theory without being especially influenced by boundary conditions outside $R$. We
can, further, envisage letting the boundary tend to infinity so that the open causaloid and the causaloid become
equivalent.

\section{Some results concerning lambda matrices}

The causaloid is either specified by giving all lambda matrices or just giving a few lambda matrices and then using
some RULES to calculate all others.  If we want to use such RULES then it will be useful to have some results
relating lambda matrices.

First we note that when omega sets multiply so do lambda matrices.
\begin{equation}\label{thensodo}
\Lambda_{l_1l_2}^{k_1k_2} = \Lambda_{l_1}^{k_1}\Lambda_{l_2}^{k_2} ~~~{\rm if}~~~
\Omega_{12}=\Omega_1\times\Omega_2
\end{equation}
where the $l_1$ might belong to any subset of the full set of allowed measurements (that are labelled by
$\alpha_1$), and likewise for $l_2$. This follows from (\ref{ifomegamultiply}) using (\ref{rlambda}) and
restricting to the given sets for the $l$'s.

Next we give the following result.
\begin{equation}\label{lambdaproduct}
\Lambda_{l_1l_2l_3}^{k_1k_2k_3}=\sum_{k'_2\in\Omega_{2\not{3}}}\Lambda_{l_1k'_2}^{k_1 k_2}
\Lambda_{l_2l_3}^{k'_2k_3} ~~~{\rm if}~~~ \Omega_{123}= \Omega_{12}\times \Omega_{\not{2} 3} ~~~{\rm and}~~~
\Omega_{23}=\Omega_{2\not{3}}\times\Omega_{\not{2} 3}
\end{equation}
where the notation $\Omega_{\not{2} 3}$ means that we form the set of all $k_3$ for which there exists
$k_2k_3\in\Omega_{23}$.  This generalizes as
\begin{equation}\label{lambdamultiproduct}
\Lambda_{l_1l_2l_3l_4}^{k_1k_2k_3k_4}=\sum_{k'_2\in\Omega_{2\not{3}}, k'_3\in\Omega_{3\not{4}}}
\Lambda_{l_1k'_2}^{k_1 k_2} \Lambda_{l_2k'_3}^{k'_2k_3} \Lambda_{l_3l_4}^{k'_3k_4} ~~~{\rm if}~~
\begin{array}{l}
\Omega_{1234}= \Omega_{12}\times \Omega_{\not{2} 3} \times \Omega_{\not{3} 4} \\
\Omega_{23}=\Omega_{2\not{3}}\times\Omega_{\not{2} 3}  \\
 \Omega_{34}=\Omega_{3\not{4}}\times\Omega_{\not{3} 4}
\end{array}
\end{equation}
and so on.   We will now prove (\ref{lambdaproduct}) (we obtain (\ref{lambdamultiproduct}) using the same proof
technique).    We have
\begin{eqnarray}
{\bf r}_{l_1l_2l_3}\cdot {\bf p} &=& {\bf r}_{l_2l_3} \cdot {\bf p}_{l_1}  \\
&=& \sum_{k'_2k_3\in\Omega_{23}} \Lambda_{l_2l_3}^{k'_2k_3} {\bf r}_{k'_2k_3}\cdot {\bf p}_{l_1} \\
&=& \sum_{k'_2k_3\in\Omega_{23}} \Lambda_{l_2l_3}^{k'_2k_3} {\bf r}_{l_1k'_2k_3}\cdot {\bf p} \\
&=& \sum_{k_1k_2\in\Omega_{12}}\sum_{k'_2k_3\in\Omega_{23}} \Lambda_{l_1k'_2}^{k_1k_2}\Lambda_{l_2l_3}^{k'_2k_3}
{\bf r}_{k_1k_2k_3}\cdot {\bf p}
\end{eqnarray}
where ${\bf p}_{l_1}$ is the state for region $R_2\cup R_3$ given that the preparation was $(X^{l_1}_{R_1}\cup
X_{R-R_1-R_2-R_3}, F^{l_1}_{R_1}\cup F_{R-R_1-R_2-R_3})$ in $R-R_2-R_3$. Also note that we have used the same
method in the last line that was used in the first three lines but for ${\bf p}_{k_3}$. Since this is true for any
${\bf p}$ we have
\begin{equation}
{\bf r}_{l_1l_2l_3} = \sum_{k_1k_2\in\Omega_{12}}\sum_{k'_2k_3\in\Omega_{23}}
\Lambda_{l_1k'_2}^{k_1k_2}\Lambda_{l_2l_3}^{k'_2k_3} {\bf r}_{k_1k_2k_3}
\end{equation}
We also have, by definition,
\begin{equation}
{\bf r}_{l_1l_2l_3} = \sum_{k_1k_2k_3\in\Omega_{123}} \Lambda_{l_1l_2l_3}^{k_1k_2k_3} {\bf r}_{k_1k_2k_3}
\end{equation}
Comparing the last two equations gives us (\ref{lambdaproduct}).

We have found some relations between lambda matrices when these lambda matrices have certain properties (such as
having omega sets which satisfy the given properties).   This proves that we do not have complete freedom to choose
lambda matrices independently of one another.  It should be possible to characterize all possible relationships
between lambda matrices so we know how much freedom we have in specifying the causaloid.  These constraints are
likely to give us deep insight into the possible nature of physical theories.

\section{Transforming lambda matrices}\label{transforming}

We write $\Lambda_{l_{\cal O}}^{k_{\cal O}}({\cal O}, \Omega_{\cal O})$.     We noted in Sec.\ \ref{causaloid} that
since the lambda matrix contains all relevant linear dependencies we can calculate (i) all other omega sets and
(ii) the lambda matrix for any other omega set.  For example, we might want to check that $\Omega'_{\cal O}$ is an
omega set and then calculate $\Lambda_{l_{\cal O}}^{k_{\cal O}}({\cal O}, \Omega'_{\cal O})$.  To do this is easy.
First we form the square matrix
\begin{equation}\label{omegaomegaprime}
\Lambda_{k'_{\cal O}}^{k_{\cal O}}({\cal O}, \Omega_{\cal O}) ~~~~ k'_{\cal O} \in \Omega'_{\cal O}
\end{equation}
If this has an inverse then $\Omega'_{\cal O}$ is an omega set.  Then it is easy to show that
\begin{equation}
\Lambda_{l_{\cal O}}^{k'_{\cal O}}({\cal O}, \Omega'_{\cal O}) = \sum_{k_{\cal O}}[\Lambda_{k'_{\cal O}}^{k_{\cal
O}}({\cal O}, \Omega_{\cal O})]^{-1} \Lambda_{l_{\cal O}}^{k_{\cal O}}({\cal O}, \Omega_{\cal O})
\end{equation}
by considering the equations by which the lambda matrices are defined.  Similar remarks apply to local lambda
matrices $\Lambda_{\alpha_x}^{k_x}(x,\Omega_x)$.

Note that if $\Omega'_{\cal O} = \Omega_{\cal O}$ then the matrix in (\ref{omegaomegaprime}) is equal the the
identity. Indeed, this is how we can deduce the omega set from the lambda matrix.

\section{Introducing time}

\subsection{Foliations}

It is common in physics to think of a state evolving in \lq\lq time".  We will show how we can recover this notion
in the causaloid formalism.  This will be a useful exercise if we wish to make contact with those physical
theories, such as QT, that take the notion of an evolving state as fundamental.  We will find, however, that this
formalism admits a much more general framework for evolving states.  In particular there is no requirement that the
time slices are space-like.

If we wish to think of the state evolving in the region $R$ then we must introduce introduce a time label $t=0, 1,
\dots T$ and a set of nested subsets of $R$
\begin{equation}\label{foliation}
  R\equiv R(0) \supset R(1)\supset \cdots \supset R(t) \supset R(t+1) \supset \cdots \supset R(T)\equiv \emptyset
\end{equation}
We will call this a {\it foliation}. It is a feature of the present approach that we are free to allocate the
nested set in this foliation any way we wish - even ways that would not correspond to our usual notion of time.
However, we expect that certain foliations will be well behaved - namely those that correspond to a good choice of
$t$.  We define an {\it elementary time-slice}
\begin{equation}
R_t=R(t+1)-R(t)
\end{equation}
The elementary time-slice $R_t$ consists of all the cards in $R$ between times $t$ and $t+1$.   Note the notational
difference between $R(t)$ and $R_t$.  We use an argument to denote what happens after time $t$ and a subscript to
denote what happens between times $t$ and $t+1$.

\subsection{States evolving in time}

We can write
\begin{equation}
{\bf p}(t) \equiv {\bf p}(R(t))
\end{equation}
for \lq\lq the state at time $t$". Given this state we can calculate probabilities for what happens at times after
$t$.  If we know the causaloid then can find omega sets $\Omega(t)\equiv\Omega_{R(t)}$. The components of the state
are $p_{k(t)}$ with $k(t)\in\Omega(t)$. The notation $k(t)$ is a little unnatural. $k_t$ would be more natural, but
we reserve this for elementary time-slices.  We understand the $t$ argument on $k(t)$ to tell us that these $k$'s
are in the omega set $\Omega(t)$.

The state will evolve from time $t$ to time $t+1$.  The transformation will depend on what was done and what was
seen in the elementary time slice $R_t$.  We denote this by $(X_t, F_t)$ and have the associated vector ${\bf
r}_{(X_t, F_t)}(R_t)$. Since $R(t+1)= R_t \cup R(t)$ we can can calculate the components of ${\bf p}(t+1)$ from
${\bf p}(t)$.
\begin{equation}
p_{k(t+1)}= {\bf r}_{k(t+1)}(t+1)\otimes^{\Lambda} {\bf r}_{(X_t, F_t)}(R_t) \cdot {\bf p}(t)
\end{equation}
where $k(t+1) \in \Omega(t+1)$.   Hence we can write,
\begin{equation}
{\bf p}(t+1) = Z_{(X_t, F_t)}(t+1, t) {\bf p}(t)
\end{equation}
where the elements of the $|\Omega(t+1)|\times|\Omega(t)|$ real matrix $Z$ are given by
\begin{equation}
Z_{(X_t, F_t)}(t+1, t)_{k(t+1)k(t)} = [{\bf r}_{k(t+1)}(t)\otimes^{\Lambda} {\bf r}_{(X_t, F_t)}(R_t)] |_{k(t)}
\end{equation}
We can calculate these from the causaloid.  We can label each $(X_t, F_t)$ in the elementary time-slice $R_t$ with
$\alpha_t$.   Thus will write the transformation matrix for time $t$ to $t+1$ as $Z_{\alpha_t}$ or as $Z_t$ if we
are suppressing the $\alpha$'s.

Since there is only one preparation for $R$ (namely that implicit in the condition $C$) we know that the state at
time $t=0$ is
\begin{equation}
{\bf p}(0) \equiv {\bf p}(R) = (p_1)
\end{equation}
where $(p_1)$ is a single component vector.   Hence, we can calculate a general state by
\begin{equation}
{\bf p}(t) = Z(t, 0) {\bf p}(0)
\end{equation}
with
\begin{equation}
Z(t, 0) = Z_tZ_{t-1}\cdots Z_0
\end{equation}
We see that the causaloid provides us with a notion of a state evolving and tells us how to evolve it.  The only
quantity left undetermined by the causaloid is the component $p_1$.  But this will cancel when we use Bayes formula
to calculate conditional probabilities in $R$ and so need not be determined.

\subsection{Obtaining lambda matrices from transformation matrices}

We can write $(X_R, F_R)$ in $R$ as the union of $(X_t, F_t)$ in $R_t$ over $t=0$ to $T-1$ or denote it with the
collection of $\alpha$ labels $\alpha_{T-1}\cdots \alpha_0$.   Then
\begin{equation}
{\rm prob}(X_R, F_R)={\rm prob}(\alpha_{T-1}\cdots\alpha_0)= {\bf r}_{\alpha_{T-1}} Z_{\alpha_{T-2}}\cdots
Z_{\alpha_0}\cdot {\bf p}(0)
\end{equation}
For notational simplicity we perform the following replacements
\begin{equation}
{\bf r}_{\alpha_{T-1}} \rightarrow Z_{\alpha_{T-1}}  ~~~~{\rm and}~~~~ Z_{\alpha_0}\cdot{\bf p}(0) \rightarrow
Z_{\alpha_0}
\end{equation}
Hence we have (suppressing most $\alpha$'s)
\begin{equation}\label{ZZZ1}
{\rm prob}(X^{\alpha_t}_t\cup X_{R-R_t}, F^{\alpha_t}_t\cup F_{R-R_t}) \equiv p_{\alpha_t}=Z_{T-1}Z_{T-2}\cdots
Z_{\alpha_t} \cdots Z_0
\end{equation}
Corresponding to the elementary region $R_t$ is a state ${\bf p}_t$ with a generalized preparation $(X_{R-R_t},
F_{R-R_t})$ in the region $R-R_t$.  Note that the generalized preparation contains a part to the past of $t$ and a
part to the future of $t+1$ (at this level the framework is similar to the time-symmetric formulation of Aharonov,
Bergmann and Lebowitz \cite{ABL}). The state ${\bf p}_t$ has components
\begin{equation}\label{ZZZk}
p_{k_t} = Z_{T-1} \cdots Z_{k_t} \cdots Z_0
\end{equation}
We can write an arbitrary $Z$ at time $t$ in terms of the linearly independent set $Z_{k_t}$ with $k_t\in\Omega_t$.
That is
\begin{equation}\label{ZrZ}
Z_{\alpha_t} = \sum_{k_t\in\Omega_t} r^{\alpha_t}_{k_t} Z_{k_t}
\end{equation}
Putting (\ref{ZZZk}) and (\ref{ZrZ}) into (\ref{ZZZ1}) gives
\begin{equation}
p_{\alpha_t} = {\bf r}_{\alpha_t} \cdot {\bf p}_t
\end{equation}
as required. This justifies the use of $r_{k_t}$ in (\ref{ZrZ}) and it justifies labelling the linearly independent
set of $Z_{k_t}$'s with $k_t\in\Omega_t$.  This means that we can write
\begin{equation}\label{ZlambdaZ}
Z_{\alpha_t} = \sum_{k_t\in\Omega_t} \Lambda_{\alpha_t}^{k_t} Z_{k_t}
\end{equation}
using (\ref{rlambda}).

Now consider adjacent elementary time-slices $R_t$ and $R_{t+1}$.  We have
\begin{equation}\label{ZZZll}
p_{l_{t+1}l_t}=Z_{T-1}Z_{T-2}\cdots Z_{l_{t+1}} Z_{l_t} \cdots Z_0 {\rm ~~~for~~~} l_{t+1}l_t\in\Omega_{t+1}\times
\Omega_t
\end{equation}
There will be a state ${\bf p}_{t+1, t}$ for the composite region made from $R_t$ and $R_{t+1}$  associated with a
generalized preparation in the region $R-R_{t+1}-R_t$. This state has components
\begin{equation}\label{ZZZkk}
p_{k_{t+1}k_t}=Z_{T-1}Z_{T-2}\cdots Z_{k_{t+1}} Z_{k_t} \cdots Z_0   {\rm ~~~with~~~} k_{t+1}k_t\in \Omega_{t+1,t}
\end{equation}
We can write
\begin{equation}\label{ZZrZZ}
Z_{l_{t+1}} Z_{l_t}=\sum_{k_{t+1}k_t\in \Omega_{t+1,t}} r_{k_{t+1}k_t}^{l_{t+1}l_t} Z_{k_{t+1}} Z_{k_t}
\end{equation}
Putting (\ref{ZZZkk}) and (\ref{ZZrZZ}) into (\ref{ZZZll}) gives
\begin{equation}
p_{l_{t+1}l_t}= {\bf r}_{l_{t+1}l_t}\cdot {\bf p}_{t+1,t}
\end{equation}
Hence we can write
\begin{equation}\label{ZZlambdaZZ}
Z_{l_{t+1}} Z_{l_t}=\sum_{k_{t+1}k_t\in \Omega_{t+1,t}} \Lambda^{k_{t+1}k_t}_{l_{t+1}l_t} Z_{k_{t+1}} Z_{k_t}
\end{equation}
using (\ref{r12lambda12}).

This method will also work for more than two sequential elementary time-slices.   In general
\begin{equation}\label{ZZZlambdaZZZ}
Z_{l_{t+\tau}}\dots Z_{l_{t+1}}Z_{l_t}=\sum_{k_{t+1}k_t\in \Omega_{t+1,t}} \Lambda^{k_{t+\tau}\dots
k_{t+1}k_t}_{l_{t+\tau}\dots l_{t+1}l_t} Z_{k_{t+\tau}}\dots Z_{k_{t+1}} Z_{k_t}
\end{equation}
for $\tau$ sequential regions $R_t$ to $R_{t+\tau}$.

If a theory provides transformation matrices $Z$'s we can use (\ref{ZlambdaZ}) and (\ref{ZZZlambdaZZZ}) to obtain
lambda matrices $\Lambda_{\alpha_t}^{k_t}$ and $\Lambda^{k_{t+\tau}\dots k_{t+1}k_t}_{l_{t+\tau}\dots l_{t+1}l_t}$.
However, this is not sufficient to fully specify the causaloid (or even the open causaloid) since (i) it does not
tell us how to calculate lambda matrices for a non-sequential set of elementary time-slices and (ii) elementary
time-slices may contain many elementary regions $R_x$ and we do not know how to obtain lambda matrices between
these.   Up till now everything we have done has been quite general.  In particular, all this works for {\it any}
choice of nested regions $R(t)$ or, equivalently, for {\it any} choice of disjoint elementary time-slices $R_t$. To
deal with (ii) we need to add spatial structure which we will deal with later. In the mean time we would like to
make progress on (i) and to that end we will introduce some extra assumptions which are true in QT and CProbT for
the natural choice of time slicing.

\subsection{Some assumptions}

We make two assumptions which happen to be true in CProbT and QT.
\begin{quote}
{\it Assumption 1:} We assume $|\Omega(t)|=K={\rm constant}$ for all $t$ except the end points $t=0$ and $t=T$
where we must have $|\Omega(0)|=|\Omega(T)|=1$.
\end{quote}
\begin{quote}
{\it Assumption 2:} We assume that $|\Omega_t| = |\Omega(t)||\Omega(t+1)|$ so that we have the maximum possible
number of linearly independent matrices $Z_{k_t}$.
\end{quote}
The first assumption follows from symmetry under time translations (except at the end points).   Both assumptions
taken together imply that $|\Omega_t|=K^2$ for $t=1$ to $T-1$ and $|\Omega_{0,T-1}|=K$ for the first and last
time-slice.  Consequently the non-end point transformation matrices are $K\times K$ (that is square matrices) and
the matrices at the end points are $1\times K$ for $R_{T-1}$ and $K\times 1$ for $R_0$ (that is they are row and
column vectors respectively).

Now consider two non-sequential time-slices $R_t$ and $R_{t'}$ with $t'>t+1$.  We have
\begin{equation}\label{ZZZlandl}
p_{l_{t'}l_t}=Z_{T-1}\cdots Z_{l_{t'}} Z(t'-1,t+1) Z_{l_t} \cdots Z_0 {\rm ~~~for~~~} l_{t'}l_t\in\Omega_{t'}\times
\Omega_t
\end{equation}
where $Z(t'-1,t+1)$ is the transformation from $t+1$ to $t'-1$.  We can define the linear operator $[Z_{l_{t'}} ?
Z_{l_t}]$ by
\begin{equation}
[Z_{l_{t'}} ? Z_{l_t}]Z(t'-1,t+1) \equiv Z_{l_{t'}} Z(t'-1,t+1) Z_{l_t}
\end{equation}
It can be shown that it follows from assumptions 1 and 2 that
\begin{equation}\label{ZqZresult}
[Z_{l_{t'}} ? Z_{l_t}] {\rm ~~~for~~~} l_{t'}l_t\in\Omega_{t'}\times \Omega_t {\rm~~~are~linearly~independent}
\end{equation}
Hence,
\begin{equation}
\Omega_{t't} = \Omega_{t'} \times \Omega_t
\end{equation}
where $\Omega_{t't}$ is the omega set for region $R_{t'}\cup R_t$.   When omega sets multiply then so do lambda
matrices (see (\ref{thensodo})).  Hence,
\begin{equation}
\Lambda_{l_{t'}l_t}^{k_{t'}k_t} = \Lambda_{l_{t'}}^{k_{t'}}\Lambda_{l_t}^{k_t}
\end{equation}
where $t'>t+1$.  This result generalizes in two respects. First it works if we replace the elementary time slices
by clumps of any number of sequential time slices.   For example
\begin{equation}
\Lambda_{l_{t'}l_tl_{t-1}}^{k_{t'}k_tk_{t-1}} = \Lambda_{l_{t'}}^{k_{t'}}\Lambda_{l_tl_{t-1}}^{k_tk_{t-1}}
\end{equation}
where $t'>t+1$.  Second, it works for more than two non-sequential clumps.  For example,
\begin{equation}
\Lambda_{l_{t''+1}l_{t''}l_{t'}l_tl_{t-1}}^{k_{t''+1}k_{t''}k_{t'}k_tk_{t-1}} =
\Lambda_{l_{t''+1}l_{t''}}^{k_{t''+1}k_{t''}}\Lambda_{l_{t'}}^{k_{t'}}\Lambda_{l_tl_{t-1}}^{k_tk_{t-1}}
\end{equation}
where $t''-1>t'>t+1$ (so we have non-sequential clumps).  This second generalization requires proving a
generalization of (\ref{ZqZresult}).

We can now summarize as follows. If we have a physical theory for which there exists a choice of nested subsets
$R(t)$ such that the above two assumptions are satisfied and if this theory provides us with transformation
matrices $Z$ then we can use the above results to calculate lambda matrices for all unions of elementary
time-slices.   Given an arbitrary such union of elementary time-slices we proceed as follows.  First we use
equation (\ref{ZZZlambdaZZZ}) to obtain, from the $Z$'s, a lambda matrix for each clump of sequential time-slices.
We then multiply these lambda matrices to get the desired lambda matrix.  We can also obtain local lambda matrices,
$\Lambda_{\alpha_t}^{k_t}$, for each elementary time-slice using (\ref{ZlambdaZ}).

\subsection{Calculating lambda matrices from more basic lambda matrices}

In the previous subsection we need to use $Z$'s to calculate lambda matrices.  We would like to leave the $Z$'s
behind since, from our point of view, they belong to a less fundamental way of thinking about this structure.  In
this subsection we will show how we can get lambda matrices for arbitrary unions of elementary time-slices
(excluding the first and the last elementary time-slice) starting only with local lambda matrices,
$\Lambda_{\alpha_t}^k$, for each $R_t$ and lambda matrices, $\Lambda^{k_{t+1}k_t}_{l_{t+1}l_t}$, for pairs of
sequential $R_t$.  But to do this we need to add one more assumption.
\begin{quote}
{\it Assumption 3:}  We assume that one at least one of the allowed transformation matrices, $Z_t$, for each
elementary time-slice (except for the first and the last), $R_t$, is invertible so $Z^{-1}_t$ exists.
\end{quote}
Note, we do not require that $Z^{-1}_t$ is in the set of allowed transformation matrices. In $R_t$ we have the
fiducial matrices $Z_{k_t}$ with $k_t\in\Omega_t$. We let
\begin{equation}\label{omegat1L}
\Omega_t = (1, 2, 3, \dots, L)  ~~~\text{for}~~~ 0<t <T-1
\end{equation}
where we have $L=K^2$.  Employing the above assumption, we can,  without loss of generality, choose the first
fiducial matrix, $Z_1$, to be invertible for each elementary time-slice except the first and last.  Now consider a
clump of sequential elementary time-slices from $R_t$ to $R_{t'-1}$ (with $t>0$ and $t'<T$) where we implement
$Z_1$ for each elementary time-slice except the first in the clump where we implement $Z_{k_t}$. The corresponding
matrices
\begin{equation}
Z_1Z_1\dots Z_1Z_{k_t} {\rm ~~~with~~~} k_t\in \Omega_t
\end{equation}
form a linearly independent set in terms of which we can expand general transformations $Z(t',t)$ from $t>0$ to
$t'<T$.  Hence, we can say that a fiducial set is given by
\begin{equation}\label{fidposs}
\Omega_{t, t+1,\cdots, t'-1} = (111\cdots1, ~211\cdots1, \cdots,~ L11\cdots1 )
\end{equation}
It is very simple to verify that these omega sets satisfy the conditions given for (\ref{lambdaproduct}) and its
generalizations (such as (\ref{lambdamultiproduct})) to hold.  Hence, using this method we can calculate the lambda
matrix for an arbitrary sequential clump {\it using just the lambda matrices for pairwise sequential regions} (so
long as we exclude the first and the last elementary time-slices). We can then apply the methods of the last
section to get the lambda matrix for a completely arbitrary set of elementary time-slices (though still excluding
the first and last elementary time-slice).

\subsection{A basic causaloid diagram}

\begin{figure*}[p!]
\resizebox{\textwidth}{!} {\includegraphics{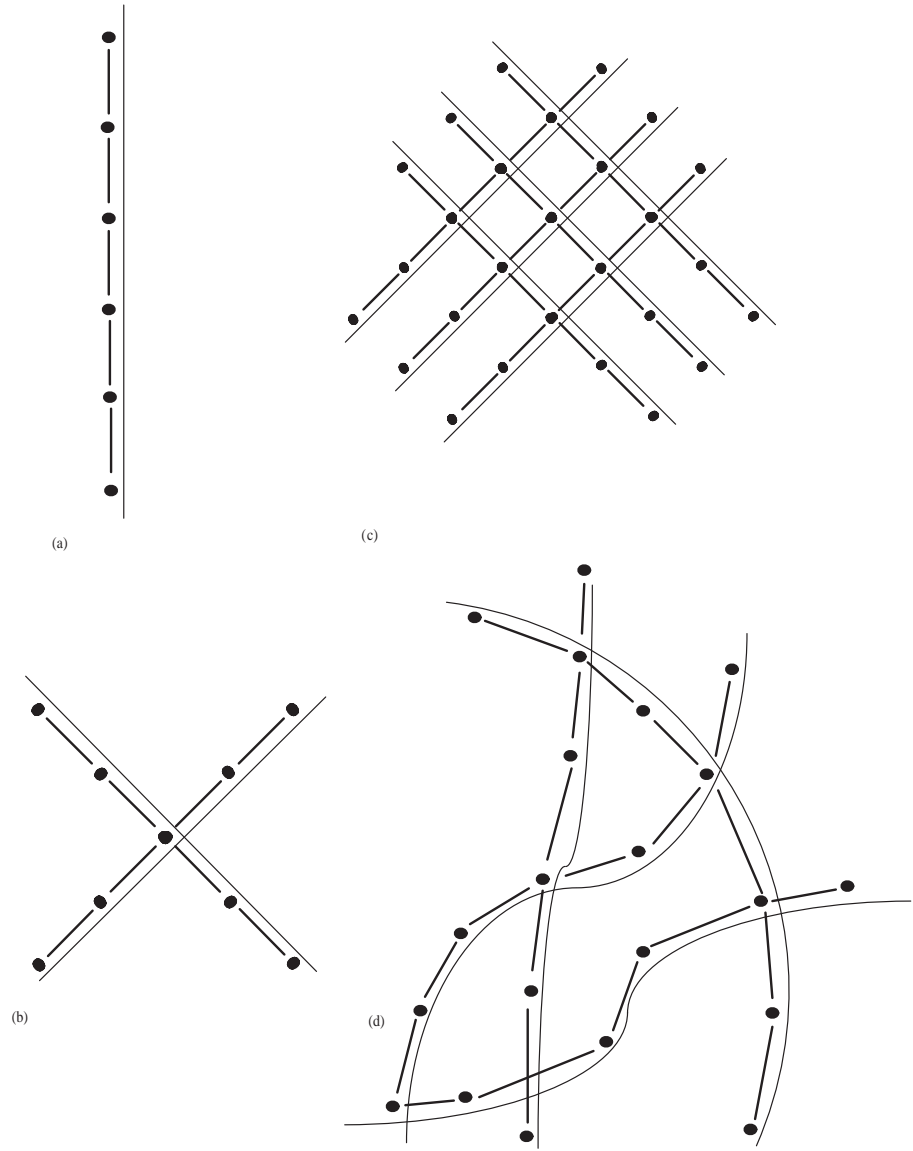}} \caption{\small A causaloid diagram for (a) a single
system (b) two interacting systems, and (c,d) a number of systems interacting.}\label{cdiagram}
\end{figure*}

Assume that the elementary time-slices are, in fact, elementary regions.  Then we have developed sufficient
machinery to calculate the open causaloid (where the first and last elementary time-slices are regarded as being in
the boundary region).   We can summarize this with a diagram, which we will call a causaloid diagram. This diagram
is shown in Fig.\ \ref{cdiagram}(a) consisting of nodes, links, and a sisterline whose significance is the
following.
\begin{enumerate}
\item {\bf Nodes} correspond to elementary regions $R_t$ and are dressed with $\Lambda_{\alpha_t}^{k_t}$.  From this lambda
matrix we can deduce the omega set $\Omega_t=(1, 2, \cdots, L)$ associated with the node.
\item {\bf Links} join sequential regions and are dressed with $\Lambda_{l_tl_{t+1}}^{k_tk_{t+1}}$.  From this lambda matrix
we can deduce the omega set $\Omega_{t,t+1}=(11, 21, \cdots L1)$ associated with the link.
\item The {\bf sisterline} (the thin line running along the side) denotes the set of omega sets
$\Omega_{t, t+1,\cdots, t+\tau} = (111\cdots1, ~211\cdots1, \cdots,~ L11\cdots1 )$.  This line is to the right as
we go up.  This corresponds to the direction implicit in this set of omega sets.
\end{enumerate}
We exclude the first and last elementary time-slices from this diagram.   Using the causaloid diagram we can
determines the open causaloid since we can calculate any lambda matrix (except those pertaining to the first and
last elementary time-slices). A lambda matrix for arbitrary ${\cal O}$ can be calculated using the clumping method
obtained above. We first identify all clumps of sequential $t$'s in ${\cal O}$.  Then we use (\ref{lambdaproduct})
and its generalizations (such as (\ref{lambdamultiproduct})) to calculate lambda matrices for each sequential
clump. Then we multiply these lambda matrices to get the lambda matrix for ${\cal O}$.  For clumps consisting of a
single member we use (\ref{elementarylambda}) before multiplying the lambda matrices.

If we are starting with a theory which is expressed in terms of transformation matrices (we regard such a theory as
less fundamental) then we can calculate the lambda matrices for nodes and links using (\ref{ZlambdaZ}) and
(\ref{ZZlambdaZZ}).  We will show how to do this for CProbT and QT in a later section. Once we have these lambda
matrices we can disregard the transformation matrix formalism and work with the causaloid formalism instead (as
long as Assumptions 1, 2, and 3 are true).

\section{Adding spatial structure}\label{introdspace}

To add spatial structure we will use the notion of interacting {\it systems}.  We will label systems with $i=1, 2,
\dots$.   We can regard the situation depicted in Fig.\ \ref{cdiagram}(a) as corresponding to a single  system. Now
consider the causaloid diagram shown in Fig.\ \ref{cdiagram}(b).  This depicts what we can regard as two systems
interacting. We label these systems $i$ and $j$.  These labels become attached to the corresponding sister lines.
We have two types of node. Nodes at crossing points (there is only one such node in Fig.\ \ref{cdiagram}(b)) and
nodes at non-crossing points. We can think of crossing points as having two systems present which may (depending on
what local procedure is carried out) be interacting.  More complicated situations involving several interacting
systems are shown in the causaloid diagrams in Fig.\ \ref{cdiagram}(c,d).   For simplicity we will restrict the
maximum number of systems in any given elementary region to two so we never have more than two systems crossing
through a node.  The methods we will present can quite easily be generalized to situations in which we relax this
constraint.

The nodes are labelled by $x$ which we think of as a space-time label.  For each system we have a sequence of $x$'s
(those picked out by the corresponding sister line).   Our intention is to find a way to go from theories (like
CProT and QT) which have transformation matrices to the causaloid formalism.  Thus, for each system we have a
sequence of matrices $Z_{x}^i$ and local omega sets $\Omega_x^i$. We now introduce the following assumption
\begin{quote}
{\it Assumption 4:} The matrices $Z_{k^i_xk^j_x}\equiv Z_{k^i_x}\otimes Z_{k^j_x}$ with
$k^i_xk^j_x\in\Omega^i_x\times\Omega^j_x$ form a complete fiducial set at any crossing node, $x$.  Here $\otimes$
is defined in the usual way (so that $(A\otimes B)(C\otimes D)=AC\otimes BD$).
\end{quote}
It follows from Assumption 4 that we can write a general transformation matrix at a crossing node as
\begin{equation}\label{ZotimesZ}
Z_{\alpha_x} = \sum_{k^i_xk^j_x\in\Omega^i_x\times\Omega^j_x} \Lambda_{\alpha_x}^{k^i_xk^j_x} Z_{k^i_x}\otimes
Z_{k^j_x}
\end{equation}
Thus, at the crossing node $x$ we have
\begin{equation}\label{omegaxij}
\Omega_x = \Omega_x^i\times\Omega_x^j
\end{equation}
This equation implies that $|\Omega_x|=|\Omega_x^i||\Omega_x^j|$.  We can interpret this to mean that when two
systems are put together the number of properties is simply the product of the number of properties of each system
- we do not lose or gain properties.   Equation (\ref{ZotimesZ}) tells about how systems interact.  If we can write
\begin{equation}
\Lambda_{\alpha_x}^{k^i_xk^j_x}=\Lambda_{\alpha^i_x}^{k^i_x}\Lambda_{\alpha^j_x}^{k^j_x}
\end{equation}
then we can write all transformation matrices at $x$ as $Z_{\alpha^i_x}\otimes Z_{\alpha^j_x}$ and consequently the
two systems actually cannot interact.  It would further follow that we can always write
\begin{equation}
(X_{R_x}, F_{R_x}) =  (X_{R^i_x}\cup X_{R^j_x}, F_{R^i_x}\cup F_{R^j_x})
\end{equation}
at $R_x$ and, hence, that we can regard anything we might do in $R_x$ as being composed of separate actions on the
two systems.  In this case we might as well regard $x,i$ and $x,j$ as corresponding to separate elementary regions
- there is no reason to have a crossing node.   Interaction at $R_x$ requires that
$\Lambda_{\alpha_x}^{k^i_xk^j_x}$ does not factorize (i.e. this is required for something interesting to be going
on at the crossing node $x$).  In particular, it means that there are some $(X_{R_x}, F_{R_x})$ at $R_x$ which
cannot be regarded as being composed of separate actions on the two systems.  It is interesting that we have a form
of interaction between two systems here even though the omega sets multiply simply because the local lambda matrix
does not factorize.

If local omega sets multiply for two systems at a crossing node then it is reasonable that omega sets for larger
regions for two systems will multiply.  Thus, we assume
\begin{quote}
{\it Assumption 5:} Omega sets $\Omega_{{\cal O}_1}^i$ and $\Omega_{{\cal O}_2}^j$ multiply where $i\not= j$ and
${\cal O}_1$ and ${\cal O}_2$ may overlap.
\end{quote}
With this final assumption in place we will be able to calculate the open causaloid.   First let us summarize.
\begin{enumerate}
\item Each {\bf non-crossing node} is dressed with $\Lambda_{\alpha_x}^{k^i_x}$.  The $i$ label is that of the sisterline
passing by this node.  From the lambda matrix we can deduce the omega set $\Omega_x^i=(1, 2, \cdots, L_i)$
associated with the node.
\item  Each {\bf crossing node} is dressed with $\Lambda_{\alpha_x}^{k^i_xk^j_x}$.  The $i$ and $j$ labels are those
of the sister lines crossing by this node.  From the lambda matrix we can deduce the omega set $\Omega_x =
\Omega_x^i\times\Omega_x^j$ associated with this node.
\item  Each {\bf link} is dressed with $\Lambda_{l^i_xl^i_{x'}}^{k^i_xk^i_{x'}}$.  The $i$ is that of the sister line
running along side this link.  The $x$ and $x'$ are those of the nodes at either end of the link.  From this lambda
matrix we can deduce the omega set $\Omega_{x,x'}^i$ associated with this link.
\item  Each {\bf sister line} is associated with a system and has a label $i$.  Associated with the sister line is a set
of omega sets
\[\Omega^i_{x, x', \dots, x''} = (111\cdots1, ~211\cdots1, \cdots,~ L_i11\cdots1 )   \]
for system $i$ here $x, x', \dots, x''$ are sequential nodes along the line which have the line running to the
right as we go along the sequence.
\end{enumerate}
Basically we have a lambda matrix for each node and each link along with some rules about how omega sets for
composite regions are formed.  From these we can calculate the lambda matrix for an arbitrary composite region
$R_{\cal O}$ as follows.
\begin{enumerate}
\item  Identify all clumps of nodes in ${\cal O}$ which are sequential along a given sister line $i$.   For each of
these clumps apply the procedure outlined in the previous section - namely applying (\ref{lambdaproduct}) and its
generalizations (such as (\ref{lambdamultiproduct})) to obtain lambda matrices for system $i$ for these clumps.
For clumps consisting of a single node we have $\Lambda_{l^i_x}^{k^i_x}=\delta_{l^i_x}^{k^i_x}$ as in
(\ref{elementarylambda})
\item  Repeat this for each sister line.
\item  Now multiply all the lambda matrices calculated in steps 1 and 2 to get the lambda matrix for $R_{\cal O}$.
\end{enumerate}
The local lambda matrices for the elementary regions are already given and hence we can calculate all lambda
matrices. This means that the causaloid diagram dressed with lambda matrices in the manner described is a way of
representing the open causaloid. To determine the open causaloid we need only specify a small subset of all the
lambda matrices. We then have the above RULES for calculating other lambda matrices.   We call these RULES the {\it
clumping method}.

In physical theories it is typical that we have symmetry such that the lambda matrix associated with equivalent
objects (non-crossing nodes, crossing nodes, or links) in the causaloid diagram could be identical.  In this case
we can represent the open causaloid by {\it just three lambda matrices} and the appropriate causaloid diagram.

\section{Time, space, and systems}

In the previous section we employed a picture of systems inhabiting space evolving in time and interacting when
they meet at the same space-time location.   This picture underpins much theoretical thought in physics. However,
from the point of view of the causaloid formalism this picture need not be regarded as fundamental.  Rather, it
provides an organizing principle which is useful to calculate the causaloid.  If we start with a suitable causaloid
then we may be able to work backwards and derive this picture.  If we regard the causaloid as fundamental then we
should contend with the possibility that this picture is derived.  Further, it may turn out that this picture is of
limited or no use in calculating some causaloids (and maybe the causaloid for QG is such an example). We should
therefore be wary of attempting to derive physical theories from this picture.

\begin{figure*}
\resizebox{\textwidth}{!} {\includegraphics{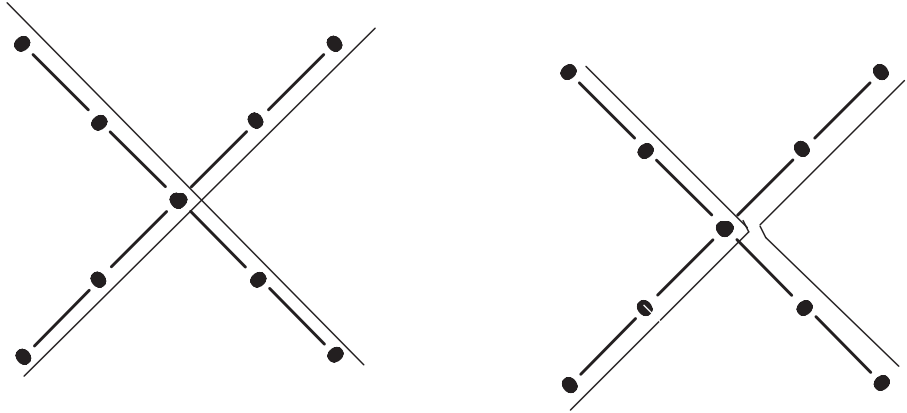}} \caption{\small It is possible that the same causaloid
can be represented by two different causaloid diagrams.  A possible example is shown in this
figure.}\label{onectwod}
\end{figure*}

Already in the above examples the causaloid can have properties which weaken the notion of system.  In particular,
we note that it it possible to represent the same causaloid by different causaloid diagrams.  We give a possible
example in Fig.\ \ref{onectwod}.

\section{The causaloid for classical probability theory}\label{classcausaloid}

We will consider a number of interacting classical bits.  An example of a bit would be a ball which can be in box 1
or box 2.

First we consider a single bit.  This has a state given by
\begin{equation}
{\bf p}(t) = \left( \begin{array}{c} p_1 \\ p_2 \end{array} \right)
\end{equation}
(see Sec.\ \ref{CProbT}).  We represent this system by a sequence of nodes labelled $t$ as in the causaloid diagram
shown in Fig.\ \ref{cdiagram}(a).  We can evolve the state by acting on  it with a sequence of transformations
$Z_{\alpha_t}$. These transformation matrices must satisfy the properties outlined in Sec.\ \ref{CProbT}.  To see
what this entails first we can write
\begin{equation}
Z_{\alpha_t} = \left(\begin{array}{cc} z^{\alpha_t}_{11} & z^{\alpha_t}_{12} \\
z^{\alpha_t}_{21} & z^{\alpha_t}_{22}  \end{array}\right)
\end{equation}
We can interpret $z^{\alpha_t}_{mn}$ as the probability that the ball jumps to box $m$ and outcome $\alpha_t$
happens  given that the ball is in box $n$.  Hence,
\begin{equation}\label{spaceofZ}
0 \leq z^{\alpha_t}_{mn} {\rm ~~~and~~~}  \sum_m z^{\alpha_t}_{mn} \leq 1
\end{equation}
For each value of the label ${\alpha_t}$ we will have a different realization of $Z$ consistent with these
constraints.  This space of allowed $Z$'s is continuous and hence the set of labels $\alpha_t$ will be infinite.
However, $\alpha_t$ is supposed to label actual data that may be recorded on a card and so must actually belong to
a finite set.  Thus, in practice we will only include a finite set of possible $Z$'s in the set we can actually
implement.  Typically recording $\alpha_t$ will involve reading numbers off scales to some accuracy and the finite
resolution in doing this will lead to a finite set.

The following allowed $Z$'s can serve as fiducial matrices
\begin{equation}
{Z}_1=\left(\begin{array}{cc} 1 & 0 \\ 0 & 1 \end{array} \right) ~~~
{Z}_2=\left(\begin{array}{cc} 1 & 0 \\ 0 & 0 \end{array} \right)~~~
{Z}_3=\left(\begin{array}{cc} 0 & 0 \\ 1 & 0 \end{array} \right)~~~
{Z}_4=\left(\begin{array}{cc} 0 & 1 \\ 0 & 0 \end{array} \right)
\end{equation}
Note that $Z_1$ is invertible satisfying Assumption 3.  In words, ${Z}_1$ is the identity transformation and leaves
the system unchanged. ${Z}_2$ is the transformation associated with a measurement to see if the ball is in box 1
which leaves the ball in box 1 afterwards. This means that in a backwards time direction it also measures to see if
the ball is in box 1. ${Z}_3$ is the transformation associated with a measurement to see if the ball is in box 1
and which flips it into box 2 if it is.  In the backwards time direction it looks to see if the ball is in box 2
and flips it into box 1. ${Z}_4$ is similar to ${Z}_3$ with the box labels interchanged.

Something interesting has happened here.  We have a classical bit and so the number of reliably distinguishable
states is $N=2$.  However, since we are in a setting where the generalized preparation of region $R_t$ is both to
the past and the future, we have $K_t\equiv |\Omega_t| = N^2$.   In the quantum case we will have $K_t\equiv
|\Omega_t| = N^4$. The point is that in this generalized preparation setting it is the transformation matrices $Z$
that map linearly to the ${\bf r}$ vectors.

We can write
\begin{equation}
Z_{\alpha_t} = \sum_{k_t = 1}^4 \Lambda_{\alpha_t}^{k_t} Z_{k_t}
\end{equation}
This equation can be solved to give the components of the local lambda matrix $\Lambda_{\alpha_t}^{k_t}$ in terms
of the matrix elements of $Z_{\alpha_t}$.
\begin{equation}\label{locLambdaforbit}
\Lambda_{\alpha_t}^1=z_{22}^{\alpha_t}, ~~\Lambda_{\alpha_t}^2=z_{11}^{\alpha_t} - z_{22}^{\alpha_t},  ~~
\Lambda_{\alpha_t}^3=z_{21}^{\alpha_t}, ~~ \Lambda_{\alpha_t}^4=z_{12}^{\alpha_t}
\end{equation}
The real importance of these expressions is given by the fact that we know the space of allowed $Z_{\alpha_t}$
(given in (\ref{spaceofZ})).  Given this we can now calculate the space of allowed $\Lambda_{\alpha_t}^{k_t}$.  The
purpose of the $\alpha_t$ label is then simply to label each element in this space.

Now consider two sequential times $t$ and $t+1$.  An omega set is given by
\begin{equation}
\Omega_{t+1, t} = (11, 12, 13, 14)
\end{equation}
We have
\begin{equation}
Z_{l'}Z_{l} = \sum_{k'k\in\Omega_{t+1,t}} \Lambda_{l'l}^{k'k} Z_{k'} Z_{k}
\end{equation}
We can solve these equations explicitly to obtain $\Lambda_{l'l}^{k'k}$.   Omitting the trivial elements
($\Lambda_{l'l}^{k'k}=\delta_{l'l}^{k'k}$ for ${l'l},{k'k}\in (11, 12, 13, 14)$) we have
\begin{equation}\label{bitLambda}
\begin{array}{cccccccccccccc}
\Lambda   & l'l & 21 & 22 & 23 & 24 & 31 & 32 & 33 & 34 & 41 & 42 & 43 & 44 \\
k'k  \\
11  &    &  0 & 0  &  0 &  0 &  0 &  0 &  0 &  1 &  0 &  0 &  0 &  0  \\
12  &    &  1 & 0  &  0 &  0 &  0 &  0 &  0 & -1 &  0 &  0 &  0 &  0  \\
13  &    &  0 & 0  &  0 &  0 &  1 &  1 &  0 &  0 &  0 &  0 &  0 &  0  \\
14  &    &  0 & 0  &  0 &  1 &  0 &  0 &  0 &  0 &  1 &  0 &  0 &  0
\end{array}
\end{equation}
We have now calculated the lambda matrices for nodes and links and hence we have enough to specify the open
causaloid.

Now consider having a number of bits interacting according to one of the causaloid diagrams in Fig.\
\ref{cdiagram}(b-d).  We switch to labelling nodes by $x$ and label different systems by different $i$. We have
already calculated lambda matrices for non-crossing nodes and links. To calculate the lambda matrix for a crossing
node we have to solve
\begin{equation}
Z_{\alpha_x} = \sum_{k^i_xk^j_x\in\Omega^i_x\times\Omega^j_x} \Lambda_{\alpha_x}^{k^i_xk^j_x} Z_{k^i_x}\otimes
Z_{k^j_x}
\end{equation}
The constraints on the space of $Z_{\alpha_x}$ (namely that elements are positive and the sum of each column is no
greater than 1) induces a constraint on the space of $\Lambda_{\alpha_x}^{k^i_xk^j_x}$ and the point of $\alpha_x$
is simply to label elements in this space.

We have explicitly calculated the lambda matrix for a non-crossing node and for a link and shown how to calculate
it for a crossing-node.  Given these three lambda matrices and the causaloid diagram we have fully specified the
open causaloid and therefore provided a complete predictive framework for CProbT.

\section{The causaloid for quantum theory}\label{meatends}

We could proceed in an exactly analogous way in QT as we did in CProbT.  Thus, we could take a number of
interacting qubits. Of course, we don't have to restrict ourselves to qubits. We could have systems whose Hilbert
space dimension is different from 2 and we could have systems of various Hilbert space dimension.

Thus, first let us consider a single system with associated Hilbert space of dimension $N$.  We have a causaloid
diagram as shown in Fig.\ \ref{cdiagram}(a). If we follow exactly the technique above then we would use Z matrices
for quantum theory. We saw in Sec.\ \ref{QTrZp} how QT can be formulated with Z matrices.  However, we can instead
proceed with superoperators, $\$ $, which are more familiar. In fact there is an invertible linear map between
superoperators and $Z$ matrices (\ref{dollarZ}) so we can switch between the two objects at any time. First we
choose a fiducial set of linearly independent superoperators for each $t$
\begin{equation}
\$_{k_t}    {\rm ~~~~ for ~~~~} k_t\in \Omega_t =(1, 2, \dots, N^4)
\end{equation}
Then, since there is a linear map between $Z$'s and $\$ $'s, we can write
\begin{equation}\label{dlambdad}
\$_\alpha = \sum_{k_t\in\Omega_t} \Lambda_{\alpha_t}^{k_t} \$_{k_t}
\end{equation}
instead (\ref{ZlambdaZ}).  Similarly we can write
\begin{equation}\label{ddlambdadd}
\$_{l_{t+1}} \$_{l_t}=\sum_{k_{t+1}k_t\in \Omega_{t+1,t}} \Lambda^{k_{t+1}k_t}_{l_{t+1}l_t} \$_{k_{t+1}} \$_{k_t}
\end{equation}
instead of (\ref{ZZlambdaZZ}) where
\begin{equation}
\Omega_{t+1,t} = (11, 12, 13, \dots, 1N^4)
\end{equation}
We can solve (\ref{dlambdad}) to find the space of $\Lambda_{\alpha_t}^{k_t}$ from the known space of the
superoperators.  Then the $\alpha_t$ label is used to label each point in this space (or at least a large set of
points consistent with the resolution of the experiment).  We can also solve (\ref{ddlambdadd}) to get the lambda
matrix for pairs of sequential time-slices (which we are taking to be elementary regions).  This matrix will be
$|\Omega_t\times \Omega_t|$ by $|\Omega_{t+1,t}|$.  That is it will be $N^8\times N^4$.  In the case of a qubit
this is $256\times 16$.  This is a rather big object (though not too big).  This size can be thought of as the
price we pay for working in a framework (the causaloid formalism) capable of expressing any physical theory (at
least any physical theory which correlates data as described in Sec.\ \ref{statement}). We now know the lambda
matrices for nodes and links and so have specified the open causaloid for this quantum system of dimension $N$.

We can now consider many quantum systems interacting as shown in Fig.\ \ref{cdiagram}(b-d). The $i$th such system
has Hilbert space dimension $N_i$. We switch to labelling nodes with $x$'s.  The above calculations provide us with
the lambda matrices for non-crossing nodes and for links.  We only need to find the lambda matrix for crossing
nodes.  This is given by solving
\begin{equation}
\$_{\alpha_x} = \sum_{k_x^ik_x^j\in\Omega_x^i\otimes\Omega_x^j}  \Lambda_{\alpha_x}^{k_x^ik_x^j}
\$_{k_x^i}\otimes\$_{k_x^j}
\end{equation}
which is obtained from (\ref{ZotimesZ}).  On solving this we obtain the space of $\Lambda_{\alpha_x}^{k_x^ik_x^j}$
from the known space of $\$ $'s for the two systems.  We label points in this space with $\alpha_x$ (up to the
resolution of the experiment).   Given this lambda matrix we now have the open causaloid and so can leave the usual
quantum formalism behind.

\section{The causaloid without boundary conditions}\label{noboundary}

The causaloid is defined for a predictively well defined region $R$ with a condition $C$ on the cards outside $R$.
The open causaloid is defined to exclude cards in the boundary region $R_{{\cal O}_b}$. The idea is that condition
$C$ is only relevant to this boundary region - conditional probabilities the remainder of $R$ are unaffected by
$C$. If we are restricting our attention to $R-R_{{\cal O}_b}$ then we might ask why we had condition $C$ in the
first place. If we are not interested in the cards that go into verifying $C$ why even collect these cards? Put
another way, can we simply identify $R-R_{{\cal O}_b}$ with the full pack $V$? If we retrace our steps we can see
that the reason we introduced $C$ was so we could have conditional probabilities of the form ${\rm Prob}(X_R,
F_R|C)$. Completely unconditional probabilities make no sense. However, although we use probabilities which are
only conditioned on $C$ as intermediate steps in our construction of the open causaloid, when we use the open
causaloid to calculate probabilities we are calculating the probability of something in $R-R_{{\cal O}_b}$
conditionalized on something else that happened in $R-R_{{\cal O}_b}$.  The conditioning on $C$ is also implicit,
but if we accept that conditional probabilities in $R-R_{{\cal O}_b}$ are independent of $C$, then this
conditioning on $C$ is actually redundant. This motivates us to now define the causaloid in the following way.
\begin{quote} {\bf The causaloid} for the full pack $V$ made up of  elementary regions $R_x$ is, if it exists,
defined to be that
thing represented by any mathematical object which can be used to obtain ${\bf r}_{\alpha_{\cal O}}(R_{\cal O})$
for all measurements $\alpha_{\cal O}$ in region $R_{\cal O}$ for all $R_{\cal O}\subseteq V$ where these ${\bf
r}$ vectors can be used to calculate conditional probabilities using (\ref{bayes}).
\end{quote}
When we say that we use (\ref{bayes}) to calculate conditional probabilities we mean that we look to the case where
the probabilities are independent of the state ${\bf p}(R_1\cup R_2)$ which basically means that we use
(\ref{bayesparallel}) or (\ref{pbounded}).

If we look back at the two cases we have explicitly worked out, CProbT and QT, then we see that we can actually
regard the open causaloid as the causaloid for the full pack $V$.  By defining this object we have effectively
removed the boundary condition and so have a much more useful object.  The causaloid for the full pack $V$ fully
characterizes the physical theory without the qualifications given in Sec.\ \ref{physicaltheories}.  However, we
cannot actually be sure that there will exist such a causaloid.  It is possible that conditions outside the region
of the experiment will always influence conditional probabilities inside $V$ and, if these are not taken into
account, we cannot have well defined probabilities in $V$ and so cannot have a causaloid for $V$.   One way to
avoid this would be to take the causaloid to correspond to the whole universe so there can be no possibility of
influences outside $V$. If we do this we get into various issues such as what it means to have repeatability and
what it means to take data for the whole universe. We will discuss these issues in Sec.\ \ref{universalc}.

\section{Dynamic causal structure}

The causaloid is a fixed object.  Yet at the same time we have not assumed any fixed causal structure in deriving
the causaloid formalism.  That is to say we have not specified any particular causal ordering between the
elementary regions.  In this sense we must have allowed the possibility of dynamic causal structure. It interesting
to see a little more explicitly how this can work in the causaloid formalism.

\begin{figure*}
\resizebox{\textwidth}{!} {\includegraphics{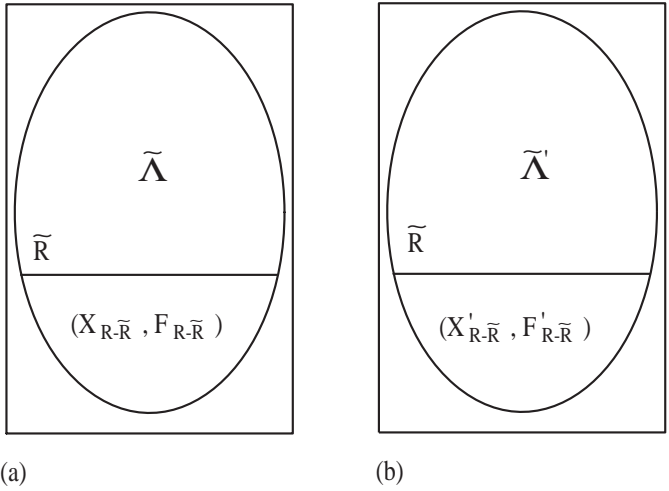}} \caption{\small Dynamic causal structure can manifest
itself when the effective causaloid in $\tilde{R}$ depends on the data in $R-\tilde{R}$.}\label{dynamic}
\end{figure*}

\begin{figure*}
\resizebox{\textwidth}{!} {\includegraphics{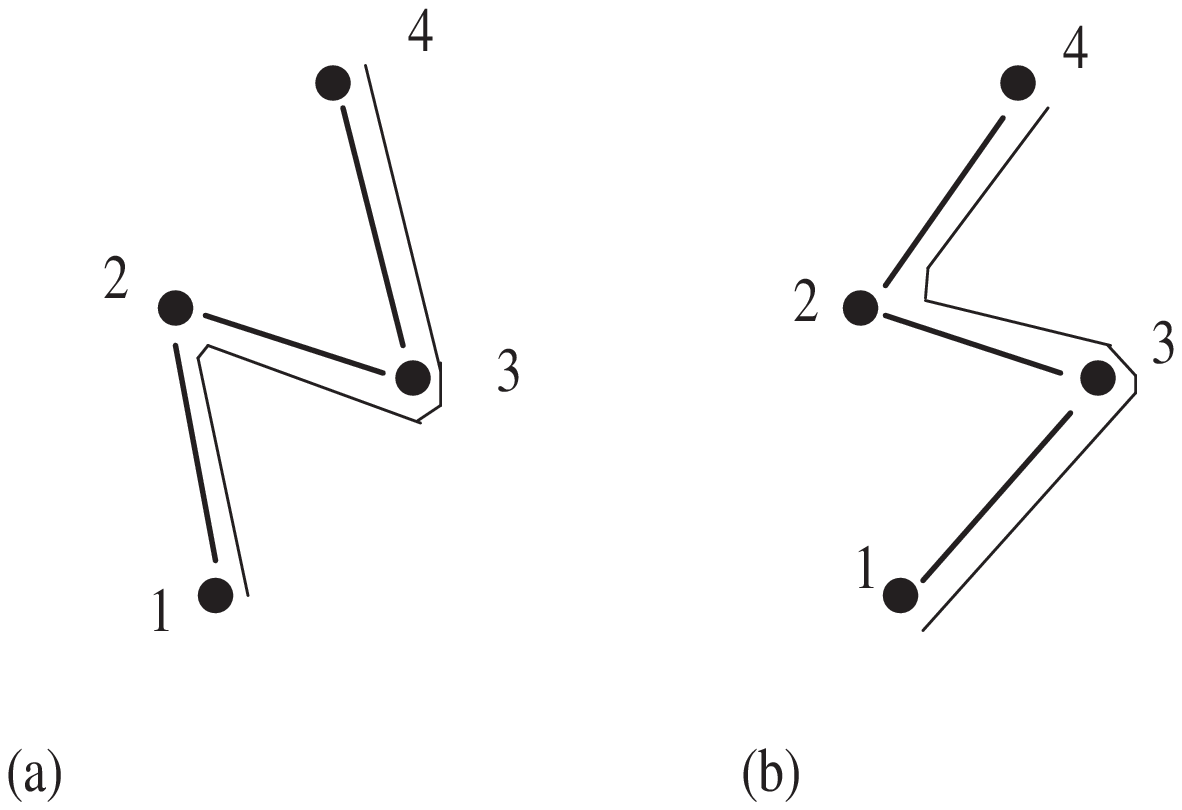}} \caption{\small Depending on data collected in some
other region, the causal structure among these four nodes may be as in (a) or (b).}\label{funnynodes}
\end{figure*}

The best way to see this is as follows.  Given a causaloid for $R$ (we could consider it for $R-R_{{\cal O}_b}$, or
$V$ instead) we can imagine that we have collected data $(X_{R-\tilde{R}}, F_{R-\tilde{R}})$ in region
$R-\tilde{R}$. We can regard this new data as conditioning, like $C$, for a new causaloid in $\tilde{R}$.  Let us
call this new causaloid $\tilde{\bf \Lambda}$.  We could alternatively imagine collecting different data,
$(X'_{R-\tilde{R}}, F'_{R-\tilde{R}})$, in the same region $R-\tilde{R}$ and obtain the causaloid
$\tilde{\Lambda}'$ for the same region $\tilde{R}$.  This is shown in Fig.\ \ref{dynamic}.   Both $\tilde{\Lambda}$
and $\tilde{\Lambda}'$ can be calculated from the original causaloid.   Now, it is possible that the causal
structure evident in $\tilde{\Lambda}$ is quite different to that evident in $\tilde{\Lambda}'$. For example, it
could be that some subset of nodes in $\tilde{R}$ has the causaloid diagram diagram shown in Fig.\
\ref{funnynodes}(a) if the causaloid is $\tilde{\Lambda}$ whereas the same nodes have the causaloid diagram shown
in Fig.\ \ref{funnynodes}(b) if the causaloid is $\tilde{\Lambda}'$ (note though  that a general causaloid cannot
be represented by causaloid diagrams of this type).

Consideration of the causaloid diagrams for CProb T and QT leads us to the conclusion that these theories do not
have dynamic causal structure of this type.  We can see that this type of reconditionalization will not change the
pattern of links.  This supports our starting assertion that CProbT and QT have fixed causal structure.

Dynamic causal structure is likely to be quite generic for causaloids.  However, it is unlikely to be as clear cut
as the hypothetical example we just discussed.  In general we cannot expect the sort of clear cut causal structure
we see evident in the causaloid diagrams of Fig.\ \ref{funnynodes}.  In general, the causal relationship between
nodes may be more complicated than can be represented by pairwise links.  Thus, when we speak of \lq\lq causal
structure" we do not necessarily intend to imply that we have well defined causal structure of the type that allows
us to determine whether two nodes are separated by a time-like or a space-like interval.

The causaloid can be thought of as containing all potential causal structures.  The particular causal structure
that gets manifested depends on what data is collected and even after data has been collected it may not be useful
or even possible to say retrospectively that two regions were time-like or space-like separated.

\section{Problems with putting general relativity in the causaloid framework}

General relativity is, ultimately an empirical theory concerning data and so it should be possible to put it in the
framework here or one very similar to it.   However,  as things stand, there are a number of obstacles.

General relativity is based on a space-time continuum whereas the structure we have described, being based on
actual data, is discrete.   There are various approaches we could take to dealing with this.  First, we could
attempt to find a continuous version of the causaloid formalism.  This would involve taking limits to pass from a
discrete to a continuous structure.  Alternatively, we could use a discrete version of GR such as the Regge
calculus \cite{regge} (for other discrete approaches see \cite{GambiniPullin}, \cite{dynamicaltriangulations}).
Since our ultimate objective is to formulate a theory of QG which is likely to be discrete and then show that GR
emerges as a limiting case we might well be satisfied with a discrete version of GR. We will discuss issues
pertaining to the continuum in a the next section.

General relativity is deterministic.  If general relativity had an external time we could use CProbT and form
probabilistic mixtures of pure states.  CProbT is well defined and so this would be a fairly straight forward
matter.  Deterministic GR would then simply correspond to the evolution of pure states in this framework. However,
since we do not have external time this option is not open to us.  Rather we have to exploit the causaloid
structure.  Despite the fact that a theory of quantum gravity would require us to address the issue of
probabilities in the context of GR, there appears to be very little work in this area.  One issue is that in GR we
tend to find a solution for a whole space-time whereas probabilities are often understood in the context of a
repeatable situation.   Even if we could create many copies of the universe, we have can only have access to data
from one copy.  This issue, at least, can be addressed (see Sec.\ \ref{universalc}).

We have introduced the notion of agency.  We allow the possibility of alternative actions - such as the setting of
knobs.  Agency is actually a very common notion in physics.  In Newtonian physics we speak of forces being applied.
In quantum theory we think of performing different measurements.   In both classical and quantum physics, we can
remove the need for agency by, for example, including a full description of the agents in the Hamiltonian. Even so,
we might ask whether we can really understand physical theories without some notion of agency. An equation in
physics tells us what would happen were various counterfactual possibilities realized.  In quantum theory the
notion of agency is especially entrenched.   The quantum state is, one might argue, best understood as a list of
probabilities pertaining to incompatible measurements - that is pertaining to different possible actions - and is
therefore difficult to interpret without agency. The notion of agency is not naturally incorporated into Einstein's
field equations. However, we could attempt to incorporate it by interpreting the eventual effects of tiny
differences in $T_{\mu\nu}$ well below the experimental resolution we are working to as corresponding to different
choices of an agent.  For example, whether a knob is set in one position or another depends on tiny differences in
the brain of the experimentalist. A special case is where there is only ever one choice of action. And consequently
no-agency is a special case of agency. This means, at least formally, there is no problem of putting GR into an
agency based framework. However, if do not have active agency we lose something potentially quite important. The
notion of causal structure is best understood from an agency point of view.  Thus, we can ask whether measurement
outcomes in one location depend on what action is implemented in another location. We cannot employ this
understanding of causality if we do not have an active notion of agency.

In GR coordinates, $x^\mu$, are usually understood to be abstract.  If follows from the principle of general
covariance that, having found a solution $(g_{\mu\nu}, T_{\mu\nu})$ on some manifold $\cal M$, the correct
description of nature is given by forming the equivalence class over all diffeomorphisms of that manifold (smooth
mappings of the points in $\cal M$ onto itself).  The abstract coordinates, $x^\mu$, themselves have no physical
meaning.  However, the $x$ which appears on the cards in the causaloid formalism is an actual observation and must
be read off an actual physical system.  In one sense this is better. The cards have actual data on them and, in
this sense, must correspond to observables.  In some sense we are (from the point of view of GR) already working
with diffeomorphism invariant objects. However, this leaves the problem of how to put GR in the causaloid
formalism. One approach to this problem is to attempt to introduce actual physical coordinates into GR.  Einstein,
in a semi-popular account of GR \cite{popEinstein}, spoke of a \lq\lq reference-mollusc" as a way of giving
physical meaning to the abstract coordinates.  Thus he imagined many small clocks attached to a non-rigid reference
body such that infinitesimally displaced clocks have readings that are infinitesimally close (see also
\cite{molluscrefs}).  We could consider many different molluscs. Einstein then states \lq\lq The general principle
of relativity requires that all these molluscs can be used as reference-bodies with equal right and equal success
in the formulation of the general laws." Another physical reference frame is provided by the fact that the universe
is, at a fine grained level, highly non-isotropic and inhomogeneous.  Therefore the view from each point is
different. This provides a way of physically labelling each space-time point (or at least each small region). A
modern version of a physical reference frame is the GPS system. Any space-time point could be labelled with the
retarded times received from four appropriately positioned clocks. Rovelli has considered how one might go about
measuring the metric using such a GPS system \cite{GPS}. General relativity has the property, in common with other
pre-quantum physical theories, that it is fairly harmless to consider counterfactuals.  Thus, even if there is no
reference mollusc, we can consider, counterfactually, what would have happened if there had been one. Rather than
having $T_{\mu\nu}$ we have $T_{\mu\nu}+ T^{\rm mollusc}_{\mu\nu}$. If $T^{\rm mollusc}_{\mu\nu}$ is sufficiently
small compared to $T_{\mu\nu}$ then the two solutions of Einstein's field equations will differ very little. Hence,
we can draw empirical conclusions about the solution with no mollusc by looking at the solution with a mollusc. In
quantum theory such counterfactuals are notoriously more tricky. Two solutions differing only in a single photon
can be quite radically different.  A successful attempt to find a theory of QG should embrace rather than shy away
from this issue.


\section{New calculus}\label{newcalculus}

Newton invented differential calculus for the purpose of understanding the motion of particles.  In doing so he
considered a ratio $\delta x / \delta t$ and took the limit as $\delta t \rightarrow 0$.  There are two ways we
might understand this limit.  We might imagine that time is ontologically continuous and so it makes sense to
consider smaller and smaller time intervals.  Alternatively, we might take an operational approach.  Thus, we might
imagine that there is no limit to how accurately we can measure $\delta x$ and $\delta t$.  If we take an
operational approach then we know that there is certainly a practical limit to how accurately we can measure these
intervals and, furthermore, there may be limits in principle.

In practice rather than directly measuring $\delta x$ and $\delta t$ to some incredible precision we often perform
a measurement over a much longer time and extrapolate back to get a value of $\delta x / \delta t$.  For example,
we may deduce the transverse velocity of a particle emerging from a hole in a card from the position, $y$, it hits
a screen placed at some distance.  We will have some model of the motion of the particles which will allow this
deduction.  Tiny differences in this transverse velocity will correspond to big differences in $y$. And this is
really the key point.  The reason we want to imagine that we have a well defined ratio $\delta x / \delta t$ is
that, even though it is essentially impossible to measure it as defined, our models predict that tiny differences
in the ratio correspond to big differences at later times.   There is a danger that the quantity $\delta x / \delta
t$ only derives it meaning from the model and this process of extrapolation and that it does not actually have
either the ontological or operational meaning we allot to it.

Even though very small differences in $\delta x / \delta t$ may not be measurable, the much bigger differences in
$y$ at the screen are.  Ideally we would like a calculus which is not based on quantities that are not directly
measurable (and may have no meaning) but is still predictively useful.  How are we to account for the measurable
differences in $y$ other than in terms of small differences in some quantity $\delta x / \delta t$ we cannot
measure?  In fact, in developing the causaloid formalism, we have already given an answer to this. We developed the
notion of fiducial measurements.  Thus, though we might be able to measure a large set of quantities, it is only
necessary to measure a small fiducial set to deduce all the others. Thus, rather than relating $y$ to a quantity
that cannot actually be measured, we can simply relate it to other quantities like $y$ which can actually be
measured.  The causaloid formalism gives a consistent way of doing this (though for probabilities).  For example,
we might imagine that we can apply various fields to particles emerging from the hole, place the screen at various
distances, and so on.  We would like to know the position the particle is detected on the screen for all these
various things we might do.  These various positions will be related and the causaloid formalism provides the
appropriate mathematical machinery for relating them.

One reason we seek to define quantities like $\delta x / \delta t$ is that we want to know the state at a given
time $t$.  The usual notion of \lq\lq state at time $t$" is that it pertains to some ontological state of affairs
at time $t$ which can be measured, in principle, at time $t$ or at least within some short time $t$ to $t+\delta
t$. But it is not necessary that nature admits such notions at a fundamental level.   And, even if nature does not
admit these notions, we will still be able to do empirical science using the causaloid formalism.  In the causaloid
formalism the basic object, $\bf \Lambda$, is built out of lambda matrices.  These matrices pertain to
operationally defined elementary regions at the level of experimental apparatuses.  We can use the causaloid to
give meaning to the state at time $t$ by choosing a foliation as in (\ref{foliation}).  The state at time $t$ is
${\bf p}(t)\equiv {\bf p}(R(t))$ where $R(t)$ contains everything of interest that comes after time $t$.   The
fiducial measurements $\Omega(t)$ are in $R(t)$.  A stronger constraint would be to demand that the fiducial
measurements actually fall in $R_t \equiv R(t+1)-R(t)$.   That is we might demand that it be possible to establish
the state at time $t$ by measurements carried out during the small time interval $t$ to $t+1$.   This requirement,
or something like it, would seem to be a feature of all current physical theories.  However, there is no reason to
demand it in principle.  If we drop this constraint then move beyond the type of situation envisaged by Newton
where the state at time $t$ is specified by quantities measurable in principle during the time interval $t$ to
$t+\delta t$. Thus, we see that the causaloid formalism provides us with a new calculus capable of dealing with
situations where Newton's differential calculus would be inappropriate.

The advantage of differential calculus and the implied ontology is that, where it works, it affords a simple
picture of reality which allows significant symmetries to be applied.  We can hope that increased familiarity with
the causaloid approach may achieve something similar.

In classical physics, including GR, the distinction between ontological notions of space and time and the
operational support for them is not an important one.  There are no fundamental limitations coming from classical
theories on how small apparatuses which might measure $\delta x$ and  $\delta t$ can be.   In QT the situation is a
little more subtle.  QT as applied to say systems of atoms does predict a scale which suggests a limit on how small
apparatuses might be. However, we can imagine that there are other fields which can probe nature on a smaller scale
and there is no limit coming from QT as to how small this scale might be.  In these cases we can always imagine in
principle probing on much smaller scales than the characteristic scales of the physics being considered. It is only
when we get to QG that we hit in principle limits to our ability to probe nature directly at smaller and smaller
scales.  Any instrument used to probe at these small scales will necessarily have mass and energy.  As $\delta x$
and $\delta t$ become smaller the associated energy and momentum will lead to black hole formation. This happens at
the Planck scale.  Thus in QG we expect for the first time a clear break down in our ability to give operational
support for ontological notions of continuous time and space (see also \cite{Ishamcontin}). We should be wary of
introducing ontological notions which are not backed up, at least in principle, by operational procedures since we
risk introducing factitious elements into our theory.  Hence, in QG we expect that we will have to use a different
calculus. The causaloid calculus provides a way forward here.

The formulation of ProbGR is likely to be useful in formulating QG.  However, it is worth noting that this likely
operational breakdown of the continuum distinguishes QG from ProbGR and constitutes an extra problem that we must
deal with.

It is often stated that experiments to test a theory of QG will involve probing nature at the Planck scale.   It is
no coincidence that apparatuses we might construct to do this would have to be very big.  As illustrated above,
postulated variation at a small scale shows up at a large scale and we might even doubt that there is any
ontological meaning to talking about what is happening on this small scale.  The fiducial measurements in the
causaloid formulation for such an experiment will, we expect, be at a much larger scale than the Planck scale.

\section{Ideas on how to formulate GR and QG in the causaloid framework}

In the causaloid framework the lambda matrices that make up the causaloid tell us everything.  In a formulation of
GR they must therefore replace both $g_{\mu\nu}$ and $T_{\mu\nu}$.   In this respect it is interesting to note that
the two roles of $g_{\mu\nu}$ pointed out in Sec.\ \ref{remarksGR} have analogues in the causaloid formalism. Thus,
the local lambda matrices tell us about local physics as does the value of $g_{\mu\nu}$ at a point, and the
composite system lambda matrices tell us about how elementary regions become correlated as does the connection
$\Gamma^\alpha_{\mu\nu}$ (which depends on the local variation of $g_{\mu\nu}$).  There are two approaches we could
take to formulating GR in the causaloid framework.  First we could attempt to put a discrete version of GR (such as
the Regge calculus) in the framework.  Secondly, we could attempt to rederive GR from scratch in this new framework
perhaps by imitating appropriate aspects of Einstein's original derivation of GR.  The second approach is likely to
be more fundamental though the first approach may help us gain important insights.

Data is collected on cards and is typically of the form $(x, a, s)$.  We can think of $x$ as playing two roles.
First, it provides a label that differentiates the elementary regions and second, it provides a local orientation
in four-dimensional space-time. The action $a$ will typically have some direction associated with it. Thus, we
might measure the spin along a certain axis. If we want think relativistically we should use a four vector $a_\mu$
to describe this measurement (this being measured relative to the local orientation provided by $x$). Similarly the
outcome $s$ will also have a direction and should be denoted with a four vector $s_\nu$. The measurement $(X_{R_x},
F_{R_x})$ would therefore be associated with two four-vectors.  It is reasonable, therefore, to suppose that a
fiducial set is given just be taking such measurements associated with the sixteen components (as happens in
tensorial analysis). Thus, we can select a subset of all measurements where $a$ and $s$ are orientated along the
$\mu$ and $\nu$ directions respectively. This will lead to fiducial measurements $k_{\mu\nu}$.  For each $\mu\nu$
there will be many such $k$'s corresponding to each possible value of $a$ and $s$. The situation may be more
complicated since $a$ and $s$ may have more $\mu$-type labels.  Motivated by the discussion in Sec.\
\ref{introdspace} we could attempt to formalize this by saying that in each elementary region $R_x$ there exist
fields $k^i_{\mu\nu}(x)$ where
\begin{equation}
\Omega(x)= \Omega^i(x)\times \Omega^{i'}(x) \times \cdots \times \Omega^{i''}(x)
\end{equation}
with
\begin{equation}
\Omega^i(x) = (\text{All}~~~ k^i_{\mu\nu}(x) ~~~ \text{for given} ~~~x, i  )
\end{equation}
Using suggestive notation we could have fields $g_{\mu\nu}$ and $T_{\mu\nu}$.  We would then have local lambda
matrices
\begin{equation}
\Lambda_{\alpha_x}^{g_{\mu\nu}(x)T_{\sigma\tau}(x)}
\end{equation}
and composite system matrices
\begin{equation}
\Lambda_{h_{\mu\nu}(x)h_{\sigma\tau}(x')}^{g_{\mu\nu}(x)g_{\sigma\tau}(x')} ~~~~\text{and}~~~~
\Lambda_{S_{\mu\nu}(x)S_{\sigma\tau}(x')}^{T_{\mu\nu}(x)T_{\sigma\tau}(x')}
\end{equation}
where we use $h$ and $S$ to label the precompression elements in the product omega sets.   We can form similar
objects for composite systems composed of more than two elementary regions.  Since we do not have fixed causal
structure we do not expect the methods of Sec.\ \ref{introdspace} to work.  If these initial steps are correct then
the problem of formulating both GR and QG in this framework is to find the corresponding causaloids.

In general relativity, the principle of general covariance requires that the form of the laws, expressed as
equations between tensorial objects, is invariant under general coordinate transformations.  We could attempt to do
something similar in the causaloid formalism.   Thus, let us imagine that the causaloid is determined by solving
some equations.   Then there are two levels at which we could demand something like general covariance. First, at a
general level.  We can write down the causaloid with respect to an arbitrary set of omega sets for each region
$R_{\cal O}$.   We could require that the equations which determine the causaloid take the same form for any choice
of omega sets.   Second, and less generally, we could just require that these equations are invariant under general
transformations of the coordinates $x^\mu \rightarrow x'^\mu$ which induces a transformation $k^i_{\mu\nu}(x)
\rightarrow k'^i_{\mu\nu}(x)$ of the local fiducial measurements, which, in turn, induces a transformation
$\Omega^i(x)\rightarrow \Omega'^i(x)$ in the local omega sets.

At least in the case of GR we should seek a way to implement the principle of equivalence.  As a first stab at this
we might require that there always exists a coordinate transformation inducing a transformation to local omega sets
$\Omega_{\rm SR}^i(x)$ such that the local lambda matrices predict special relativistic physics in $R_x$.  However,
since we are in a probabilistic context we have to admit the possibility that we have uncertainty as to what this
local frame is.  A sufficiently precise measurement in $R_x$ should be able to establish what the set of local
inertial frames is for each field $i$.   We can let this measurement be associated with a set of measurement
vectors ${\bf r}^i_{g_{\mu\nu}}$ where $g_{\mu\nu}$ labels the local inertial frames which leave $g_{\mu\nu}$ in
Minkowski form.   The equivalence principle requires that the fields are correlated so that, when there is
certainty as to what the local inertial frame is for each field, there is agreement.  Thus,
\begin{equation}
\text{If    } {\bf r}^i_{g_\mu\nu}(x)\cdot{\bf p}(x) = 1 \text{  and   } {\bf r}^{j}_{h_\mu\nu}(x)\cdot{\bf p}(x)=1
\text{    then   } h_{\mu\nu}=g_{\mu\nu}
\end{equation}
This should work in GR.  However, in QG we expect to get pure states which are superpositions of other pure states.
Thus, if we have ${\bf p}_1(x)$ and ${\bf p}_2(x)$ each of which agrees with special relativistic physics but with
different $g_{\mu\nu}$, then we expect other pure states corresponding to some sort of superposition of these for
which there is no transformation to special relativistic physics.  The implementation of the equivalence principle
in QG would therefore seem to require a preferred basis with respect to which it holds.

If we are successful in formulating GR (actually ProbGR) in the causaloid framework we can then attempt to
formulate QG.  It is quite likely that some of the differences between ProbGR and QG mirror the differences between
CProbT and QT and this might give us a strong handle on how to obtain QG from ProbGR.  There are two key
differences between CProbT and QT.

First, in CProbT we have $K=N$ whereas in QT we have $K=N^2$ (in the sense discussed in Sec.\ \ref{QTpostulates}).
Thus, if we are to build up a complete set of fiducial measurements in QG it is likely that we will want to add
extra measurements to those GR.  In fact the situation is a little more complicated for the causaloid than it was
in Sec.\ \ref{QTpostulates}.  Since the preparation for the elementary region $R_x$ is both to the future and the
past (and the sides) of $R_x$ it is the transformation matrices which map linearly to the ${\bf r}$ vectors and
hence, as discussed in Sec.\ \ref{classcausaloid}, we have $|\Omega_x|=N^2$ in CProbT and $|\Omega_x|=N^4$ in QT.
To get from to QT from CProbT we need to add two new fiducial ${\bf r}$'s for each pair of fiducial measurements in
CProbT.

The second key difference is that QT has a continuous set of reversible transformations whereas CProbT has only a
discrete set. This has the effect of filling out the space of pure states in QT. In QT we have unitary
transformations. This is unlikely to survive in QG since we do not have a fixed background to evolve the state with
respect to. Thus, information about the state is likely to leak out into the degrees of freedom which represent our
frame of reference.   For a transformation to be reversible on the other hand we require that no information about
the state leaks out. However, we may have something which very well approximates reversible unitary transformation
for sufficiently small regions of space time.

In the above discussion we used $g_{\mu\nu}$ and $T_{\mu\nu}$.  However, operational quantities would be better
than these.  Operational quantities would involve the meeting of test particles, the behaviour of rays of light,
and so on \cite{EPS, Woodhouse, BaezBunn}.

\section{The universal causaloid}\label{universalc}

We have defined a few notions of causaloid: (i) the causaloid for a predictively well defined region $R$; (ii) the
open causaloid; and (iii) the causaloid for the whole pack $V$ (this may not actually exist).  We now wish to
introduce a further notion - the universal causaloid.   The motivation for this is to remove the need for
repeatability.  We repeat an experiment many times and bundle the cards from each run together forming a stack. The
fact that we are able to bundle the cards separately indicates that, actually, there is some additional marker
which could constitute recorded data that distinguishes the cards from one run to the next. For example, in the
case of the probes floating in space illustrated in Fig.\ \ref{probes} we reset the clocks and take note of the
fact that we have done this.  All this is rather artificial. Why should we bundle our data into stacks coming from
different repetitions of an experiment? It would, one might imagine, be more natural simply to collect one big
stack of data which might or might not be regarded as coming from repetitions of an experiment. The problem with
this is that it is unclear how we might interpret probabilities. The following approach seems reasonable.
Associated with any proposition $A$ concerning the data that might be collected is some vector ${\bf r}_A$.   In
testing the data to see whether $A$ is true we will be testing its truth among a complete set of mutually exclusive
propositions $A, A', \dots, A''$. We define ${\bf r}^I_A = {\bf r}_A + {\bf r}_{A'} + \cdots + {\bf r}_{A''}$.   We
will say that $A$ is true if
\begin{equation}\label{rAapproxrAI}
{\bf r}_A \approx {\bf r}_A^I
\end{equation}
we use the symbol $\approx$ because we can never expect experimental data to give absolute support for a
proposition.  We can decide in advance just how exactly equal we require these two vectors to be.

To illustrate how this can apply to the case of probabilities consider the vectors
\begin{equation}
{\bf r}_n \equiv {\bf r}_{(X_{1n}, F_{1n})}\otimes^\Lambda{\bf r}_{(X_{2n}, F_{2n})}
\end{equation}
pertaining to in the disjoint regions $R_n\equiv R_{1n}\cup R_{2n}$ for $n=1$ to $N$ where $N$ is big. Further,
define
\begin{equation}
{\bf r}_n^I \equiv \sum_{Y_{1n}\subseteq F_1} {\bf r}_{(Y_{1n}, F_{1n})}\otimes^\Lambda{\bf r}_{(X_{2n}, F_{2n})}
\end{equation}
We define $\overline{\bf r}_n = {\bf r}_n^I - {\bf r}_n$.  Now assume that
\begin{equation}\label{rprI}
{\bf r}_n = p \, {\bf r}_n^I
\end{equation}
for all $n$.  We see that ${\bf r}_n$ is like ${\bf v}$ and ${\bf r}_n^I$ is like ${\bf u}$ of Sec.\
\ref{cpredictions}.   In our previous language we would say that this means that the probability of $X_{1n}$ in
$R_{1n}$ given see $X_{2n}$ in $R_{2n}$ and we perform procedure $F_{1n}\cup F_{2n}$ in $R_n$ is $p$.   But we can
turn this into a statement in the form of (\ref{rAapproxrAI}). Thus, consider
\begin{equation}
{\bf r}_A \equiv \sum_{(p-\Delta p) N < |S| < (p+\Delta p)N} \left({\bigotimes_{n\in S}}^\Lambda {\bf
r}_n\right)\otimes^\Lambda \left({\bigotimes_{n\in \overline{S}}}^\Lambda \overline{\bf r}_n\right)
\end{equation}
This is the vector corresponding to the property that $pN$ out of the $N$ regions $R_n$ have outcome $X_{R_n}$ to
within $\pm \Delta p N$. We also have
\begin{equation}
{\bf r}_A^I \equiv {\bigotimes_n}^\Lambda {\bf r}_n^I
\end{equation}
Using (\ref{rprI}) we obtain
\begin{equation}
{\bf r}_A = \left(\sum_{(p-\Delta p) N < |S| < (p+\Delta p)N}  [p^n (1-p)^{N-n}] \right) {\bf r}_A^I
\end{equation}
From the theory of binomial distributions we have
\begin{equation}
\sum_{(p-\Delta p) N < |S| < (p+\Delta p)N} [p^n (1-p)^{N-n}] \approx 1 - O({1}/[\Delta p{\sqrt{N}]})
\end{equation}
Hence, if $N$ is sufficiently large, condition (\ref{rAapproxrAI}) is satisfied and we can say that the proposition
is true.

This means that we do not need to make repeatability intrinsic to the definition of the causaloid. Rather, we can
simply define a universal causaloid so that we can look for properties that are true (to within some small error).
We define
\begin{quote}
{\bf The universal causaloid} for a region made up of elementary regions $R_x$ is, if it exists, defined to be that
thing represented by any mathematical object which can be used to calculate vectors ${\bf r}_A$ for any proposition
$A$ concerning the data  collected in these elementary regions such that if the proposition is true (to within some
small error) we have ${\bf r}_A\approx {\bf r}_A^I$ where ${\bf r}_A^I= {\bf r}_A + {\bf r}_{A'} + \cdots + {\bf
r}_{A''}$ and $A, A', \dots, A''$ is a complete set of mutually exclusive propositions.
\end{quote}
We see that using this object we can recover the notion that we have probabilities by using the argument above in
reverse (though see \cite{cavesschack} for a cautionary tale on this subject).   However, the universal causaloid
is potentially a richer object. We can formulate many questions about the data as \lq\lq is proposition $A$ true?"
pertaining to situations where we have not repeated the same experiment many times.  The universal causaloid should
enable us to answer all such questions.

One problem with the universal causaloid is that we cannot directly measure it (unlike the earlier causaloids we
defined) since we do not have repeatability.   It is repeatability that allows us to obtain probabilities for
different procedures and hence calculate the lambda matrices.  However, we can suppose that there are certain
symmetries and deduce the causaloid that way.   The causaloid for CProbT and QT will simply be that found by
allowing a causaloid diagram such as that in Fig.\ \ref{cdiagram}(c) to extend indefinitely to include possible
repetitions of the experiment.  We may imagine that the causaloid extends indefinitely into the future and
arbitrarily far into the past. Indeed, we might think of the universal causaloid as corresponding to the entire
history of the universe (this would be essential if we want to consider cosmology).  However, we have the problem
that we cannot expect to collect cards from such an arbitrarily large region and send them to a sealed room.  The
universal causaloid, as a mathematical object, transcends the limited domain in which the causaloid was first
conceived.  This can be regarded as removing some of the operational scaffolding we had originally erected to help
find a more general probability calculus. Having removed this scaffolding we are still able to use the universal
causaloid to make predictions.

\section{The principle of counterfactual indifference}

As we develop the causaloid formalism we should look for simplifying assumptions.  One possible such assumption is
the following
\begin{quote}
{\bf The principle of counterfactual indifference} states that the probability of $E$ does not depend on what
action would have been implemented had $E'$ happened instead if we condition on cases where  $E'$ did not happen
(as long as the device implementing this action is low key).
\end{quote}
For example, imagine Alice tosses a dice then a coin.  Then the probability that a coin comes up heads cannot
depend on the fact that had a six come up she intended to bend the coin in a particular way if we only consider
those cases where a six did not come up. Indeed, in a different procedure she might have intended not bend the coin
had the six come up. If the principle of counterfactual indifference were false in this case, then somebody could
deduce Alice's intention from data in which a six never comes up and where she never implements her intention. But
this would contradict the principle of indifference to data since such intentions are part of the programming and
correspond to the detail of the way in which information is stored (in this case Alice's intentions as manifested
in her brain). And indeed a little thought shows that the principle of counterfactual indifference is a consequence
of the principle of indifference to data (given that low key physical systems are used to process data).  The
principle of counterfactual indifference implies
\begin{equation}
{\bf r}_{(X_1, F_1)} = {\bf r}_{(X_1, F'_1)} ~~~~{\rm where} ~~~~ X_1\subseteq F_1, F'_1
\end{equation}
since both procedures $F_1$ and $F'_1$ amount to doing the same thing if we see $X_1$.  We do, indeed, have this
property in CProbT and QT.

\section{Comparison with other approaches to QG}

The approach outlined in this paper aims at finding a framework for a theory of QG.  As such, it is quite possible
that other approaches to QG will fit within this framework.   However, there are two key aspects of the causaloid
framework which should be compared with other approaches.
\begin{enumerate}
\item We deal with data that may be collected in actual experiments.  The approach here is \lq\lq top down" rather than
\lq\lq bottom up".  All other approaches to QG start with some ideas about the structure of space and time at very
small scales (usually the Planck scale) and then attempt to build up.
\item The causaloid formalism is more general than quantum theory.  We can attempt to treat QT and GR in an even handed way
rather than requiring GR to fit fully in the framework of QT.  Most other approaches to QG take the basic form of
QT unchanged.
\end{enumerate}
The two main approaches to QG are string theory \cite{strings} and quantum loop gravity \cite{Rovellibook,
Thiemannbook, Smolininvitation} though there are other approaches.

String theory assumes a fixed non-dynamical background space-time and attempts to obtain a perturbative version of
quantum gravity (along with the other fundamental forces).   It is difficult to see how, when we go beyond the
pertubative domain, this approach could have truly dynamic causal structure as it must.  The basic picture appears
to be that of unitary evolution in standard quantum theory.  Thus, string theory fits in the usual quantum
framework.

Quantum loop gravity is canonical approach.  A canonical formulation of GR is quantized.  Thus, it is fundamentally
a $3+1$ approach and this appears to be the conceptual origin of some of the mathematical problems faced by the
program.  In treating space and time on a different footing we break the elegance of Einstein's fundamentally
covariant approach (this seems to be ok for canonical QG if we adopt the Newton-Cartan approach where $c\rightarrow
\infty$ and we have natural $3+1$ splitting \cite{Joycartan}). Rather than forcing GR into a canonical framework it
seems more natural to require that QT be put in a manifestly covariant framework. This appears to be problematic
since the notion of a state across space evolving in time is basic to the usual formulation of QT. However, the
causaloid framework allows us to go to a more fundamental formalism in which we do not have a state evolving in
time.

A new approach emerging from the theory of quantum loop gravity is the spin-foam approach (see \cite{Fotini} for a
short review). In this approach spin-foams represent four dimensional histories in space time.  The evolution
between two times is represented by an amplitude weighted sum over such spin-foams.  In this approach we see graphs
dressed with matrices.  However, the causaloid diagrams are quite different since they are fixed for a given
theory.  The notion of a history is clearly better than that of a state at a given time so far as providing a
manifestly covariant formulation is concerned.  However, a history is a rather big thing. It involves all possible
events between the two times of interest.  The causaloid formalism deals with matrices between elementary regions.
In the case that there exist RULES we may only need to specify local lambda matrices and lambda matrices for pairs
of regions (as in QT).  This is closer to Einstein's original approach than providing an amplitude for an entire
history is.

Another approach is the causal set approach of Sorkin \cite{Sorkin}.  In this approach the fundamental notion is of
points with causal relations between them.  Causal sets are taken to obey certain axioms so that they form
partially ordered sets. These sets are represented by a graph with nodes joined by links.  The partially ordered
sets of the causal set approach are actually very different to the causaloid for various reasons.  First, these
partially ordered sets are supposed to provide a picture of space-time at the Planck scale. Second, the links are
not dressed with matrices and so causal relationship between the points is not as rich as that between elementary
regions in the causaloid formalism.  And third, a given causal set is meant to provide one possible history (rather
like a spin-foam) whereas the causaloid is a fixed object (though one which contains a way of calculating
probabilities for all histories).

A more recent approach is due to Lloyd \cite{Lloyd}.  He suggests that quantum gravity should be modelled by a
quantum computation.  In so doing he is able to implement dynamic causal structure.  However, since the whole
process is embedded in standard unitary quantum theory, there is still a background time.

Another approach is non-commutative geometry pioneered by Connes \cite{Connes}.  The basic idea is that operators
like $\hat x$, $\hat y$, and $\hat z$, at a point do not commute.   This appears to fit within the framework of QT
as we have defined it (since we said nothing about commutation relations).  In standard QT the operators $A\otimes
I$ and $I\otimes B$ will commute by virtue of the way the tensor product is defined.  In the causaloid formalism
the analogous objects will not, in general, commute if $\otimes$ is replaced by $\otimes^\Lambda$.  In this sense
there may be a connection between the causaloid formalism and non-commutative geometry.

One problem which is common to most approaches which start with a Planck scale picture is that it is difficult to
account for the four dimensional appearance of our world at a macroscopic level (Smolin calls this the \lq\lq
inverse problem" \cite{Smolininverse}). Since the approach in this paper starts at the macroscopic level, it may
allow us to circumvent this problem in the same way Einstein does in GR. Thus, we would not attempt to prove that
space-time is four dimensional at the macroscopic level but put this in by hand.  This is not an option in Planck
scale approaches to QG because the constraint that a four dimensional world emerges at the macroscopic scale has no
obvious expression at the Plank scale.

The best approach, however, may be to combine an approach which posits some properties at a Planck scale with the
causaloid approach.  By working in both directions we might hope to constrain the theory in enough different ways
that it becomes unique.

\section{Conclusions}

We have developed a framework for probabilistic theories which allow dynamic causal structure.   Central to this is
an object called the causaloid.   This object is theory specific and we have calculated the causaloid for classical
probability theory and quantum theory.   We have not calculated it for GR though we presented some preliminary
ideas.   The results in this paper suggest the following program for finding a theory of QG.
\begin{enumerate}
\item Formulate probabilistic general relativity (ProbGR) in the causaloid framework.   This will involve finding
RULES to calculate the causaloid from a basic set of lambda matrices
\item Construct ${\bf r}_{\alpha_x}$ for quantum gravity from the ${\bf r}_{\alpha_x}$ of ProbGR in such a way that
we go from $|\Omega_x|=N^2$ to $|\Omega_x|=N^4$.
\item Find RULES for QG from the RULES for ProbGR.
\end{enumerate}
A particular issue we will have to pay attention to is that in ProbGR we have continuous space-time whereas we do
not expect to be able to give operational support to this notion in QG.  This may force us to be more radical in
the construction of QG than is suggested by the above three steps.

The causaloid formalism contains some elements in common with the Aharonov, Bergmann, and Lebowitz (ABL)
time-symmetric formulation of QT \cite {ABL}.  Indeed, it may even be regarded as a radical generalization of the
ABL formulation.  The ABL approach has led to a number of fascinating results where naive reasoning leads to
counterintuitive though correct results. For example, if a particle is tunnelling through a potential barrier then,
when it is in the \lq\lq forbidden" region, one might naively reason that its kinetic energy (total energy minus
potential energy) is negative (even though it should always be positive). It turns out \cite{negKE} that a certain
type of measurement of the kinetic energy (called a {\it weak measurement}) will actually give negative readings if
the state of the particle is preselected in its half evolved state and postselected in the forbidden region (using
the ABL approach).  The causaloid formalism might be expected to put such counterintuitive properties in an even
more general setting and this may contribute to our understanding of them.

The approach taken here attempts to combine the early operational philosophy of Einstein as applied to GR  with the
operationalism of Bohr as applied to QT (see \cite{Landsman} for a discussion of how Einstein and Bohr might have
engaged in a more constructive debate). We do this primarily for methodological reasons to obtain a mathematical
framework which might be suitable for a theory of QG without committing ourselves to operationalism as a philosophy
of physics. In fact it is interesting just how close this early philosophy of Einstein is to the later philosophy
of Bohr. Einstein said
\begin{quote}
The law of causality has not the significance of a statement as to the world of experience, except when {\it
observable facts} ultimately appear as causes and effects \cite {Einstein}.
\end{quote}
and
\begin{quote}
All our space-time verifications invariably amount to a determination of space-time coincidences. (\dots) Moreover,
the results of our measurings are nothing but verifications of such meetings of the material points of our
measuring instruments with other material points, coincidences between the hands of a clock and points on the clock
dial, and observed point-events happening at the same place and the same time \cite{Einstein}.
\end{quote}
Bohr said
\begin{quote}
However far the phenomena transcend the scope of classical physical explanation, the account of all evidence must
be expressed in classical terms.  (\dots) The argument is simply that by the word {\it experiment} we refer to a
situation where we can tell others what we have done and what we have learned and that, therefore, the account of
the experimental arrangements and of the results of the observations must be expressed in unambiguous language with
suitable application of the terminology of classical physics \cite{Bohr}.
\end{quote}
While Einstein might have felt uncomfortable with the lack of ontological clarity of Bohr's interpretation, there
is a striking similarity between these sentiments.  This underlines the power of operationalism as a methodology.

The formalism here was developed specifically in the hope of finding a theory of QG.  However it may find
application in other areas, even outside physics.  Indeed it may be useful in any situation where there is reason
to believe that a straightforward analysis in terms of a state evolving through time is inadequate.  An example
might be where we are trying to model the behaviour of a system that is better able to predict the future than we
are. The behaviour of such a system would appear to depend on the future in ways we could not account for in a
purely forward in time way and the causaloid formalism might be useful here. One possible example of a system of
this nature would be the financial markets.

\vspace{6mm}

\noindent{\Large\bf Acknowledgements}

\vspace{6mm}

I am grateful to Mike Lazaridis for providing inspiration for this work, and  to Jon Barrett, Joy Christian, Chris
Fuchs, Matt Leifer, Vanessa Shokeir, Lee Smolin, Rafael Sorkin, Rob Spekkens, Antony Valentini, and Hans Westman
for discussions.


\begin{thebibliography}{00}
\bibitem{Weinberg} S.\ Weinberg, {\it The quantum theory of fields} Vol.\ 1 (Cambridge University Press, 1995).
\bibitem{hardy} L.\ Hardy, {\it Quantum theory from five reasonable axioms}, quant-ph/0101012 (2001).
\bibitem{Einstein}  A.\ Einstein {\it Die Grundlage der allgemeinen Relativit\"atstheorie}, Annalen der Physik, 49,
(1916). English translation in {\it The principle of relativity}, translated by W.\ Perrett and G.\ B.\ Jeffery,
(Dover Publications, 1952).
\bibitem{Hardywheelerpaper} L.\ Hardy, {\it Why is nature described by quantum theory?}, pg.\ 45 in {\it Science
and ultimate reality}, eds. J.\ D.\ Barrow, P.\ C.\ W.\ Davies and C.\ L.\ Harper, Jr.\ ( Cambridge University
Press, 2004).
\bibitem{ADM} R.\ Arnowitt, S.\ Deser, and C.\ W.\ Misner, {\it The dynamics of general relativity}, pg. 227 in
{\it Gravitation: an introduction to current research}, ed.\ L.\ Witten (Wiley, New York, 1962).
\bibitem{Ashtakar} A.\ Ashtekar, {\it New variables for classical and quantum gravity}, Phys.\ Rev.\ Lett.\ {\bf
57}, 2244 (1986).
\bibitem{Ishamtime} C.\ J.\ Isham, {\it Canonical quantum gravity and the problem of time}, qr-qc/9210011 (1992).
\bibitem{UnruhWald} W.\ G.\ Unruh and R.\ M.\ Wald, {\it Time and the interpretation of canonical quantum gravity},
Phys.\ Rev.\ D {\bf 40}, 2598 (1989).
\bibitem{Mandle} X.\ Y.\ Zou, L.\ J.\ Wang, and L.\ Mandel, {\it Induced coherence and indistinguishability in
optical interference}, Phys.\ Rev.\ Lett.\ {\bf 67}, 318
(1991).
\bibitem{Penrose} R.\ Penrose, {\it The road to reality}, (see pg.\ 856) (Jonathan Cape, London, 2004).
\bibitem{expmtPenrose} W.\ Marshall, C.\ Simon, R.\ Penrose, D.\ Bouwmeester, {\it Towards quantum superpositions of a
mirror}, Phys.\ Rev.\ Lett.\ {\bf 91}, 13 (2003).
\bibitem{GPS} C.\ Roveli, {\it GPS observables in general relativity}, Phys.\ Rev.\ D {\bf 65}, 044017 (2002).
\bibitem{popEinstein} A.\ Einstein {\it Relativity} translated by R.\ W.\ Lawson, (Routledge Classics, London and
New York, 2001).
\bibitem{ABL} Y.\ Aharonov, P.\ G.\ Bergmann and J.\ L.\ Lebowitz, {\it Time symmetry in the quantum process of
measurement}, Phys.\ Rev.\ {\bf B134}, 1410 (1964).
\bibitem{regge} T.\ Regge, {\it General relativity without coordinates}, Nuovo Cimento {\bf 19}, 558 (1961).
\bibitem{GambiniPullin} R.\ Gambini and J.\ Pullin, {\it Canonical quantization of general relativity in discrete
space-times}, Phys.\ Rev.\ Lett.\ {\bf 90} (2003).
\bibitem{dynamicaltriangulations} J.\ Ambjorn, Z.\ Burda, J.\ Jurkiewicz, and C.\ F.\ Kristjansen, {\it Quantum
gravity represented as dynamical triangulations}, Acta Phys. Polon. B {\bf 23}, 991 (1992).
\bibitem{molluscrefs} J.\ D.\ Brown and D.\ Marolf, {\it Relativistic material reference systems}, Phys.\ Rev.\ D
{\bf 53}, 1835 (1996).
\bibitem{Ishamcontin} C.\ J.\ Isham, {\it Some reflections on the status of conventional quantum theory when
applied to quantum gravity}, quant-ph/0206090 (2002).
\bibitem{EPS} J.\ Ehlers, F.\ A.\ E.\ Pirani, and A.\ Shild, {\it The geometry of free fall and light propogation} in
{\it General relativity, papers in honour of J.\ L.\ Syng} (Oxford University Press, 1972).
\bibitem{Woodhouse} N.\ M.\ J.\ Woodhouse, {\it The differentiable and causal structures of space-time}, J.\ Math.\ Phys.\
{\bf 14}, 495 (1973).
\bibitem{BaezBunn} J.\ C.\ Baez and E.\ F.\ Bunn, {The meaning of Einstein's Equation}, qr-qc/0103044 (2001).
\bibitem{cavesschack} C.\ M.\ Caves and R.\ Schack, {\it Properties of the frequency operator do not imply the
probability postulate}, quant-ph/04090144 (2004).
\bibitem{strings} J.\ Polchinski, {\it String theory, vols.} 1 and 2, Cambridge University Press 1998.
\bibitem{Rovellibook} C.\ Rovelli, {\it Quantum gravity} (Cambridge University Press, 2004).
\bibitem{Thiemannbook} T.\ Thiemann, {\it Lectures on loop quantum gravity}, Lecture Notes in Physics 541 (2003).
\bibitem{Smolininvitation} L.\ Smolin, {\it An invitation to loop quantum gravity}, hep-th/0408048 (2004).
\bibitem{Joycartan} J.\ Christian, {\it Exactly solvable sector of quantum gravity}, Phys.\ Rev.\ D {\bf 56}, 4844
(1997).
\bibitem{Fotini} F.\ Markopoulou {\it Planck-scale models of the universe} pg.\ 550 in {\it Science
and ultimate reality}, eds. J.\ D.\ Barrow, P.\ C.\ W.\ Davies and C.\ L.\ Harper, Jr.\ ( Cambridge University
Press, 2004).
\bibitem{Sorkin} R.\ D.\ Sorkin {\it Causal sets: Discrete gravity}, qr-qc/0309009, 2003.
\bibitem{Lloyd} S.\ Lloyd, {\it The computational universe: quantum gravity from quantum computation},
quant-ph/0501135 (2005).
\bibitem{Connes}  A.\ Connes, {\it Non-commutative geometry} (Accademic Press, New York 1994).
\bibitem{Smolininverse} L.\ Smolin, {\it The case for background independence}, hep-th/0507235 (2005).
\bibitem{negKE} Y.\ Aharonov, S.\ Popescu, D.\ Rohrlich, and L.\ Vaidman, {\it Measurements, errors, and negative
kinetic energy}, Phys.\ Rev.\ A {\bf 48}, 4084 (1993).
\bibitem{Landsman} N.\ P.\ Landsman, {\it When champions meet: Rethinking the Bohr-Einstein debate},
quant-ph/0507220 (2005).
\bibitem{Bohr} N.\ Bohr, {\it Discussion with Einstein on epistemological problems in atomic physics} pg. 201 of {Albert
Einstein: Philospher-Scientist} Ed. P.\ A.\ Schilpp.  La Salle: Open Court.
\end{thebibliography}
\end{document}